\documentclass[aps,prd,11pt,notitlepage,longbibliography,nofootinbib,tightenlines,preprintnumbers,superscriptaddress]{revtex4-2}

\usepackage{booktabs}
\usepackage{amsmath,amssymb,amsfonts,mathrsfs}
\usepackage{graphicx}
\usepackage{color}
\usepackage{tikz}
\usetikzlibrary{calc}
 \usetikzlibrary{decorations.text}
 \usetikzlibrary{decorations.pathreplacing}
 \usetikzlibrary{shapes}

\usepackage{dsfont}
\usepackage[pdftex]{hyperref}
\hypersetup{colorlinks=true, linkcolor=darkred, citecolor=blue, linktoc=page}
\definecolor{darkred}{rgb}{0.8,0.1,0.1}

\makeatletter\def\l@subsubsection#1#2{}%
\makeatother

\makeatletter
\def\p@subsection{}
\makeatother

\renewcommand{\thesection}{\arabic{section}}

\numberwithin{equation}{section}
\renewcommand\theequation{\arabic{section}.\arabic{equation}} 

\usepackage{subfigure}

\def\cA{{\cal A}}

\def\p{{\partial}}

\def\CC{\ensuremath{\mathds C}}
\def\RR{\ensuremath{\mathds R}}
\def\ZZ{\ensuremath{\mathds Z}}

\DeclareMathOperator{\Vol}{Vol}

\def\Im{\mathop{\rm Im}}
\def\Re{\mathop{\rm Re}}

\definecolor{3dcolor}{rgb}{0.96,0.89,0.76}
\definecolor{4dcolor}{rgb}{0.812,0.851,0.914}

\newcommand{\nocontentsline}[3]{}
\newcommand{\tocless}[2]{\bgroup\let\addcontentsline=\nocontentsline#1{#2}\egroup}

\begin{document}

\title{Splitting interfaces in 4d $\mathcal N=4$ SYM}

\author{Christoph F.~Uhlemann} 
\email{uhlemann@maths.ox.ac.uk}

\affiliation{Mathematical Institute, University of Oxford, \\
	Andrew-Wiles Building,  Woodstock Road, Oxford, OX2 6GG, UK\\[1mm]}

\author{Mianqi Wang} 
\email{mqwang@utexas.edu}

\affiliation{University of Texas, Austin, Physics Department, Austin, TX, 78712, USA}

\begin{abstract}
We discuss entanglement entropies in 4d interface CFTs based on 4d $\mathcal N=4$ SYM coupled to 3d $\mathcal N=4$ degrees of freedom localized on an interface. Focusing on the entanglement between the two half spaces to either side of the interface, we show that applying the Ryu-Takayanagi prescription in general leads to multiple natural entanglement entropies. We interpret the different entropies as corresponding to different ways of assigning the 3d degrees of freedom localized on the interface to the two half spaces. We contrast these findings with recent discussions of universal relations for entanglement entropies in 2d interface CFTs and formulate generalized relations for 4d interface CFTs which incorporate our results.
\end{abstract}

\maketitle
\tableofcontents

\parskip 1mm

\section{Introduction and Summary}

The study of boundaries and interfaces in conformal field theories on the one hand and of quantum entanglement on the other both have long and fruitful histories, which are not disconnected. Entanglement entropy (EE) encodes central charges in even dimensions, the sphere free energy in odd dimensions, and plays an important role in the study of renormalization group monotones.
It also provides natural quantities associated with interfaces, i.e.\ the interface entropy arising as difference between the EE of a region including the interface and a region of the same shape without interface \cite{Affleck:1991tk,Friedan:2003yc,Casini:2016fgb,Jensen:2015swa,Casini:2023kyj}.
Regions ending on the interface provide an additional diagnostic: it was found in \cite{2005JPhA...38.4327P,Sakai:2008tt} for 2d CFTs that the interface even changes the leading divergence in the EE, such that the logarithmically divergent term is not fixed in terms of the 2d central charge and depends on the interface instead.
Recently, for 2d interface CFTs relations between the EEs of a half space and of an interval ending on the interface were observed in \cite{Karch:2021qhd,Karch:2022vot} and conjectured to be universal.

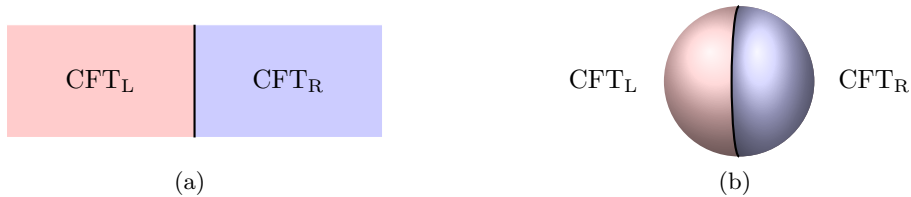
\begin{figure}
	\subfigure[][]{\label{fig:iCFT-planar}
	\begin{tikzpicture}
		\draw[white,fill=red!20!white] (0,0) rectangle (-2.5,1.5);
		\draw[white,fill=blue!20!white] (0,0) rectangle (2.5,1.5);
		\draw[thick] (0,0) -- (0,1.5);
		\node at (-1.25,0.75) {CFT$_{\rm L}$};
		\node at (1.25,0.75) {CFT$_{\rm R}$};
		
		\node at (0,-0.15) {};
	\end{tikzpicture}
	}
	\hskip 20mm
	\subfigure[][]{\label{fig:iCFT-cylinder}
	\begin{tikzpicture}[rotate=-90]
		\shade[ball color=red!20!white,opacity=1] (0,0) circle (1cm);
		
		\shade[ball color=blue!20!white,opacity=1] (1cm,0) arc (0:-180:1cm and 1mm) arc (180:0:1cm and 1cm);
		\draw[thick] (1,0) arc (0:-180:1cm and 1mm);
		\node at (0,-1.8cm) {CFT$_{\rm L}$};
		\node at (0,1.8cm) {CFT$_{\rm R}$};
	\end{tikzpicture}
	}
	\caption{Ambient CFT$_{\rm L}$ and CFT$_{\rm R}$ separated by an interface, for a planar interface in flat space on the left and an $\RR\times S^2$ interface in a 4d CFT on the cylinder $\RR\times S^3$ on the right. In our examples CFT$_{\rm L/R}$ are identical and both 4d $SU(N)$ $\mathcal N=4$ SYM. The interface hosts 3d $\mathcal N=4$ degrees of freedom. 
	We consider entanglement entropies for decompositions into subsystems $A$, $\overline{A}$ where $A\supseteq{\rm CFT}_{\rm L}$ and $\overline{A}\supseteq{\rm CFT}_{\rm R}$.
	For the cylinder geometry the EE's contain scheme-independent logarithmic terms.
		\label{fig:iCFT}}
\end{figure}

In this work we study 4d interface CFTs (ICFTs) constructed from 4d $\mathcal N=4$ SYM coupled to 3d $\mathcal N=4$ degrees of freedom on an interface in such a way that 3d defect superconformal symmetry is preserved \cite{Gaiotto:2008sa,Gaiotto:2008sd}. This is a broad space of theories, with interfaces which can in particular host genuinely lower-dimensional degrees of freedom. It can be studied efficiently using the holographic duals constructed in \cite{DHoker:2007zhm,DHoker:2007hhe}.
We realize a variety of interfaces with qualitatively different features holographically and study entanglement entropies in the ICFTs.
Motivated by the proposal for universal relations in \cite{Karch:2021qhd,Karch:2022vot}, we study the perhaps simplest EE: the entropy associated with a partition of the ICFT into the two halves to the left and to the right of the interface (fig.~\ref{fig:iCFT}). This will turn out to be non-trivial. We note that this is a different EE compared to the interface EE mentioned above, which is defined from regions which fully enclose the interface.

One may suspect the picture in fig.~\ref{fig:iCFT} to require further data, beyond the specification of the half spaces, to properly define a decomposition into two subsystems when the interface hosts additional, genuinely lower-dimensional degrees of freedom: one may expect some freedom in the choice of how the interface degrees of freedom are assigned to the two half spaces. One may then expect the EE to depend on such choices. Our results indicate that this is indeed the case. In the holographic duals for the ICFTs we discuss Ryu-Takayanagi (RT) minimal surfaces anchored on the interface. These correspond to EEs associated with decompositions of the ICFT into two subsectors, with each subsector containing one of the ambient CFT half spaces to either side of the interface. Depending on the nature of the interface we either find a unique such surface or multiple ones.

The holographic duals for 4d $\mathcal N=4$ SYM coupled to 3d $\mathcal N=4$ interfaces have 10d geometries which will be discussed in the main part.
The 3d interface degrees of freedom are 3d $\mathcal N=4$ SCFTs which are themselves holographic. The 3d holographic duals can be obtained as limiting case of the ICFT duals with vanishing number of 4d degrees of freedom. 
The qualitative form of the ICFT duals is illustrated in fig.~\ref{fig:iCFT-hol-0-a} (see also fig.~\ref{fig:D5NS5-eff} below): the 3d degrees of freedom create a large AdS$_4$ geometry, which is connected to two asymptotic ${\rm AdS}_5/\ZZ_2$ regions created by the 4d ambient CFT degrees of freedom to the left and to the right of the interface. Around each of the loci where the AdS$_5$ throats connect to the bulge representing the 3d degrees of freedom we expect a minimal surface separating the two asymptotic ${\rm AdS}_5/\ZZ_2$ regions. 
As we will discuss in more detail in the main part, the different minimal surfaces satisfy different asymptotic boundary conditions and are physically distinct; as such they compute different ICFT quantities.
So we have two candidate RT surfaces computing the EE between the two ambient CFT half spaces.
Fig.~\ref{fig:iCFT-hol-0-b} illustrates the same features from the perspective of a bottom-up braneworld type model (with geometry (\ref{eq:5d-metric1})).

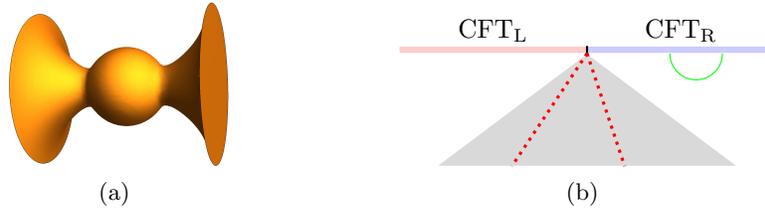
\begin{figure}
	\subfigure[][]{\label{fig:iCFT-hol-0-a}
		\includegraphics[width=0.18\linewidth]{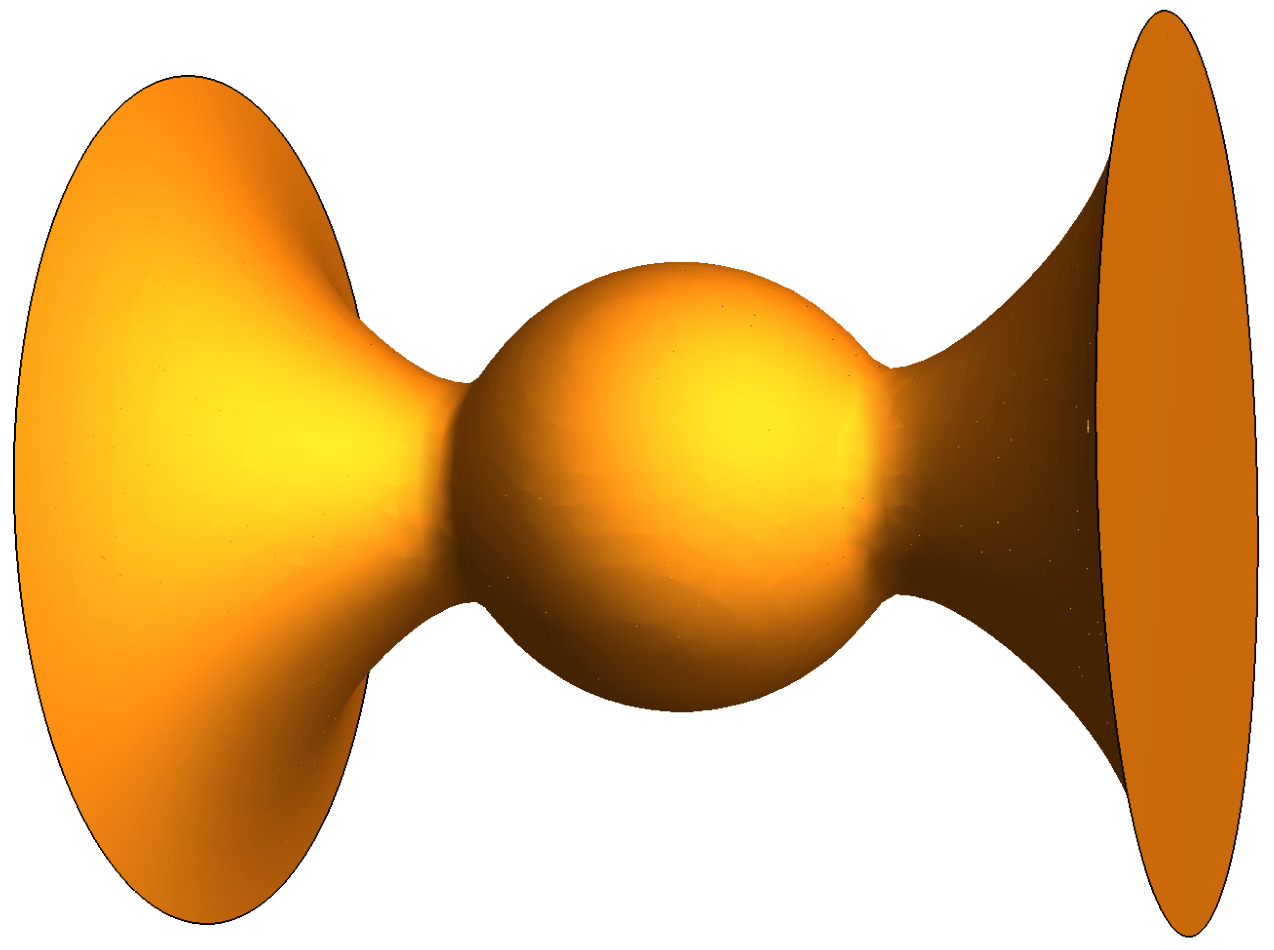}
	}\hskip 20mm
	\subfigure[][]{\label{fig:iCFT-hol-0-b}
		\begin{tikzpicture}
			\draw[white,fill=red!20!white] (0,0) rectangle (-2.5,0.1);
			\draw[white,fill=blue!20!white] (0,0) rectangle (2.5,0.1);
			\draw[thick] (0,0) -- (0,0.1);
			\node at (-1.25,0.3) {CFT$_{\rm L}$};
			\node at (1.25,0.3) {CFT$_{\rm R}$};
			
			\draw [white,fill=gray,opacity=0.3] (0,0) -- (-2,-1.5) -- (2,-1.5)--(0,0);
			\draw[very thick,dotted,red] (0,0) -- (-1,-1.5);
			\draw[very thick,dotted,red] (0,0) -- (0.5,-1.5);
			
			\draw[green] (1.8,0) arc (0:-180:0.35);
		\end{tikzpicture}
	}
	\caption{{\it Schematic} illustration of the holographic duals for CFT$_{\rm L}$ and CFT$_{\rm R}$ separated by an interface. 
		Left: Cartoon of the 10d geometries.  The AdS$_4$ geometry created by the 3d degrees of freedom is represented as a ball, with two asymptotic ${\rm AdS}_5/\ZZ_2$ regions for CFT$_{\rm L/R}$ attached as throats (compare fig.~\ref{fig:D5NS5-eff}). 
		Right: Bottom-up representation of the geometry, with the gray shaded region representing the dual to the interface degrees of freedom. Outside the shaded region the geometry approaches AdS$_5$. There may be multiple minimal surfaces anchored on the interface, represented as dotted red lines. The green line represents a surface computing the EE for an interval in one of the half spaces.
		\label{fig:iCFT-hol-0}}
\end{figure}

Depending on the features of the 3d $\mathcal N=4$ SCFT, the 3d duals themselves may permit minimal surfaces in the internal space of their duals. This was used in \cite[sec.~6]{Uhlemann:2021nhu} to realize a top-down version of wedge holography, and discussed specifically in connection with EE: the minimal surfaces in the internal space were argued to compute entanglement entropies associated with splitting the 3d SCFT into subsectors. Geometrically, one may think of the 3d duals as comprising multiple large holographic AdS$_4$ spacetimes connected by narrow bridges. 
For related discussions see \cite{Bachas:2018zmb,DeLuca:2021ojx}. 
Using such 3d $\mathcal N=4$ SCFT with minimal surfaces in the internal space as interface degrees of freedom in the 4d ICFTs leads to ICFT duals with more than two minimal surfaces associated with half space EE's (see fig.~\ref{fig:D52NS52-e} below).
We emphasize that our focus is on actual minimal surfaces, as opposed to surfaces which are merely extremal, to allow for a natural entropy interpretation.

How might this be interpreted in field theory terms? Exercising the often practiced freedom of ignoring subtleties such as gauge invariance when discussing EE, we may interpret the results in terms of tensor decompositions of the Hilbert space.
We write the ICFT Hilbert space as
\begin{align}\label{eq:H-ICFT}
	\mathcal H&=\mathcal H_{\rm CFT_L}\otimes \mathcal H_{\rm CFT_R}\otimes \mathcal H_{3d}~,
\end{align}
where $\mathcal H_{\rm CFT_L/R}$ denote the left/right ambient CFT Hilbert spaces and $\mathcal H_{3d}$ captures genuinely lower-dimensional degrees of freedom on the interface, i.e.\ fields which do not arise as restriction of ambient CFT fields to the interface. For simple interfaces $\mathcal H_{3d}$ may be trivial. 
Examples will be discussed in sec.~\ref{sec:examples}.
We then take a decomposition of the interface Hilbert space as 
\begin{align}
	\mathcal H_{3d}&=\mathcal H_{3d,A}\otimes \mathcal H_{3d,\bar{A}}~.
\end{align}
We may take $\mathcal H_{3d,A}=\mathcal H_{3d}$ with $\mathcal H_{3d,\bar{A}}$ empty, or $\mathcal H_{3d,A}$ empty with $\mathcal H_{3d,\bar{A}}=\mathcal H_{3d}$.
If the 3d degrees of freedom can be further decomposed, $\mathcal H_{3d,A}$ and $\mathcal H_{3d,\bar{A}}$ may both be non-trivial. For an interface with no 3d degrees of freedom $\mathcal H_{3d,A}$ and $\mathcal H_{3d,\bar{A}}$ may both be trivial. 
For a given decomposition of $\mathcal H_{3d}$, a decomposition of the entire interface CFT Hilbert space can then be defined as
\begin{align}
	\mathcal H_A&=\mathcal H_{\rm CFT_L}\otimes \mathcal H_{3d,A}~,
	&\mathcal H_{\bar A}&=\mathcal H_{\rm CFT_R}\otimes \mathcal H_{3d,\bar{A}}~.
\end{align}
We would expect the associated EE to depend on the decomposition of $\mathcal H_{3d}$, and if multiple choices are available we may expect multiple entropies. Our results are in line with this expectation.

The results lead to an interesting perspective on the proposal for universal relations between EE's  in 2d ICFTs \cite{Karch:2021qhd,Karch:2022vot}. The proposal relates a linear combination of the logarithmically divergent terms in the EE's for a 2d half space and in the EE for an interval ending on the interface to the central charge. An assumption in the derivation is that there is only one bulk minimal surface separating the two half spaces. 
Here we generalize this derivation to 4d ICFTs. This entails a careful choice of the geometry, as in 4d ICFTs universal logarithmic terms arise for $\RR\times S^2$ interfaces in ICFTs on a cylinder $\RR\times S^3$, but not for a planar interface in flat space.
We derive relations analogous to the 2d proposals for 4d ICFTs, but highlight that, as discussed above, for 4d ICFTs with genuine interface degrees of freedom we may have not just one but several minimal surfaces separating the two ambient half space CFTs. We generalize the EE relations to account for this fact. From that perspective the 4d versions of the 2d relations in \cite{Karch:2021qhd,Karch:2022vot} appear as a special case tied to ``simple" interfaces.
It would be interesting to explore analogous situations in 2d.

\textbf{Outline:} In sec.~\ref{sec:interfaces} we review the brane engineering and holographic duals for ICFTs based on 4d $\mathcal N\,{=}\,4$ SYM coupled to 3d $\mathcal N\,{=}\,4$ SCFT degrees of freedom on an interface, before discussing minimal surfaces in the holographic duals as well as the interface geometry from a general perspective. In sec.~\ref{sec:examples} we study concrete examples of 4d ICFTs with and without interface degrees of freedom and with varying numbers of minimal surfaces splitting the interface. 
The main features and interpretation are discussed in sec.~\ref{sec:D5NS5}.
In sec.~\ref{sec:universal} we derive EE relations for ``simple" 4d ICFTs which are analogous to the relations for 2d interfaces proposed in \cite{Karch:2021qhd,Karch:2022vot}. We discuss the implications of having multiple minimal surfaces splitting the interface and generalize the relations accordingly.

\section{Interfaces in 4d $\mathcal N=4$ SYM}\label{sec:interfaces}

\begin{figure}
	\begin{tikzpicture}[y={(0cm,1cm)}, x={(0.707cm,0.707cm)}, z={(1cm,0cm)}, scale=1.1]
	\draw[gray,fill=gray!100,rotate around={-45:(0,0,2)}] (0,0,2) ellipse (1.8pt and 3.5pt);
	
	\foreach \i in {-0.05,0,0.05}{ \draw[thick] (0,-1,2+\i) -- (0,1,2+\i);}

	\foreach \i in {-0.075,-0.025,0.025,0.075}{ \draw (-1.1,\i,2) -- (1.1,\i,2);}
	
	\foreach \i in {-0.045,-0.015,0.015,0.045}{ \draw (0,1.4*\i,0) -- (0,1.4*\i,2+\i);}
	
	\foreach \i in  {-0.075,-0.045,-0.015,0.015,0.045,0.075}{ \draw (0,1.4*\i,2+\i) -- (0,1.4*\i,4);}
	
	\node at (-0.18,-0.18,3.4) {\scriptsize D3};
	\node at (1.0,0.3,2) {\scriptsize D5};
	\node at (0,-1.25,2) {\footnotesize NS5};
	\node at (-0.18,-0.18,0.6) {{\scriptsize D3}};
	\end{tikzpicture}
	\caption{Two groups of semi-infinite D3-branes, each engineering 4d $\mathcal N=4$ SYM on a half space with a priori independent couplings. They end on groups of D5 and NS5 branes from opposite sides. The 5-brane numbers and how precisely the D3-branes terminate on the 5-branes determines the nature of the interface.	
		\label{fig:iCFT-brane}}
\end{figure}
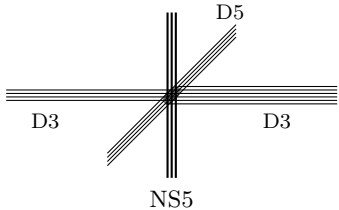

We consider supersymmetric interfaces in 4d $\mathcal N=4$ SYM engineered by D3-branes ending on and/or intersecting D5 and NS5 branes (fig.~\ref{fig:iCFT-brane}) \cite{Gaiotto:2008sa,Gaiotto:2008sd}.
The D3-branes extend along the (0123) directions, the D5-branes along (012456), and the NS5-branes along (012789).
We will focus on concrete examples in the following, and refer to the references for more systematic discussions.

The simplest interfaces join two ambient 4d $\mathcal N=4$ SYM theories on half spaces which differ in a marginal coupling. Non-trivial boundary conditions for (part of) the 4d $\mathcal N=4$ SYM fields can be implemented by terminating some D3-branes on D5-branes, and additional 3d degrees of freedom localized on the interface can be realized by NS5 branes. The general interface corresponds to a combination of D5 and NS5 branes engineering a 3d $\mathcal N=4$ SCFT through a Hanany-Witten brane setup \cite{Hanany:1996ie}, which is coupled at both ends to ambient 4d degrees of freedom on half spaces.

\subsection{Holographic duals}

The supergravity solutions corresponding to the brane configurations discussed above were constructed in \cite{DHoker:2007zhm,DHoker:2007hhe} and elaborated upon in \cite{Aharony:2011yc,Assel:2011xz}. The geometry is a warped product of the form ${\rm AdS_4}\times S^2\times S^2\times\Sigma$ with a Riemann surface $\Sigma$. 
We will only need Einstein-frame metric
\begin{align}\label{eq:metric-10d}
	ds^2&=f_4^2 ds^2_{\rm AdS_4}+f_1^2 ds^2_{S_1^2}+f_2^2 ds^2_{S_2^2}+4\rho^2 |dz|^2~, 
\end{align}
where $z$ is a complex coordinate on $\Sigma$. The line elements are $ds^2_{\rm AdS_4}$ for unit radius AdS$_4$ and $ds^2_{S_{1/2}^2}$ for two $S^2$'s. The warp factors are given by
\begin{align}\label{eq:10d-warp-factors}
	f_4^8&=16\frac{N_1N_2}{W^2}~, & f_1^8&=16h_1^8\frac{N_2 W^2}{N_1^3}~, & f_2^8&=16 h_2^8 \frac{N_1 W^2}{N_2^3}~,
	&
	\rho^8&=\frac{N_1N_2W^2}{h_1^4h_2^4}~,
\end{align}
with
\begin{align}
	W&=\partial\bar\partial (h_1 h_2)~, & N_i &=2h_1 h_2 |\partial h_i|^2 -h_i^2 W~.
\end{align}
The ${\rm AdS}_4$ in the geometry represents the defect conformal symmetry, and the boundary of the ${\rm AdS}_4$ fibers is the location of the interface. The solutions preserve 16 supersymmetries and the two $S^2$'s represent the $SU(2)\times SU(2)$ R-symmetry.

Concrete solutions are specified by a choice of $\Sigma$ and a pair of harmonic functions $h_{1/2}$ on $\Sigma$. For the solutions of interest here, $\Sigma$ can be taken as a strip with complex coordinate $z=x+iy$,
\begin{align}
\Sigma&=\left\lbrace z=x+iy\in\CC \ \big\vert \ 0\leq \Im(z)\leq \frac{\pi}{2}\right\rbrace~.
\end{align}
The general picture is that (locally) ${\rm AdS_5}\times S^5$ regions emerge at $\Re(z)\rightarrow \pm\infty$. 
Each asymptotic ${\rm AdS_5}\times S^5$ region contributes a half space as conformal boundary; representing the two half-space ambient 4d $\mathcal N=4$ SYM theories.
One of the $S^2$'s collapses on each boundary of $\Sigma$, to form a closed internal space.
The differentials $\partial h_{1/2}$ have poles on the upper/lower boundary of $\Sigma$, which correspond to D5/NS5 branes. The locations and residues encode the 5-brane charges and how the D3-branes end on or intersect the 5-branes. If 5-branes are present, the region around the sources represents the interface degrees of freedom, corresponding to the bulge/gray shaded region in fig.~\ref{fig:iCFT-hol-0}.

\subsection{Surfaces splitting $\Sigma$}

To compute the EE associated with a decomposition of the ICFT into two half spaces we seek codimension-2 minimal surfaces in the 10d geometries which wrap AdS$_4\times S_1^2\times S_2^2$ on a constant time slice,\footnote{The setups have no time dependence and can be continued to Euclidean signature; the more general HRT construction \cite{Hubeny:2007xt} reduces to the RT prescription on a constant time slice.} but split $\Sigma$, so as to separate the two asymptotic ${\rm AdS_5}\times S^5$ regions at $\Re(z)\rightarrow \pm\infty$. 
We will restrict our discussion to surfaces which preserve the symmetries of a constant time slice of $\rm AdS_4$ and $S^2\times S^2$.
Such surfaces are defined by the curve $\gamma$ along which they extend on $\Sigma$,\footnote{Similar surfaces were discussed in \cite{Uhlemann:2021nhu} for holographic duals of 3d SCFTs. The ``island surfaces" encountered there do not appear here due to the presence of the asymptotic ${\rm AdS}_5\times S^5$ regions.}
\begin{align}
	\gamma:\quad z=x_c(y)+iy~, \qquad 0\le &y\le \frac{\pi}{2}~.
\end{align}
The area of such a surface, resulting from the 10d metric in (\ref{eq:metric-10d}), is
\begin{align}\label{area-0}
 A_\gamma&=V_{{\rm AdS}_4,t={\rm const}}V_{S_1^2\times S_2^2}S_{\gamma}~,
 &
 S_\gamma&=2\int dyf_1^2f_2^2f_4^3\rho\sqrt{1+x'(y)^2}~.
\end{align}
For later convenience we express $S_\gamma$ as follows, using the expressions in (\ref{eq:10d-warp-factors}),
\begin{align}\label{area-1}
	S_\gamma&=32\int dy \sqrt{f(x,y)}\sqrt{1+x'^2}~,
&
	f&=-(h_1h_2)^3W~.
\end{align}
The boundary conditions which have to be imposed where the surface ends on $\partial\Sigma$ follow from the discussion in \cite{Uhlemann:2021nhu}: One of the $S^2$'s collapses on each boundary, and the 8d minimal surface has to close off smoothly. The curve has to end in such a way that the induced metric on the surface does not have a conical singularity.
This leads to the Neumann boundary conditions
\begin{align}\label{eq:Neumann-bc}
	x_c'(0)&=x_c'(\frac{\pi}{2})=0~.
\end{align}
The goal is to solve for the curve $\gamma$ satisfying these boundary conditions and the extremality condition
\begin{equation}
	\frac{\partial S_\gamma}{\p x}=\p_y\left(\frac{\p S_\gamma}{\p(\p_yx)}\right)\,.
\end{equation}
In terms of $f$ defined in (\ref{area-1}) it becomes
\begin{equation}
	\p_xf-x_c'\p_yf-2f\frac{x_c''}{1+x_c'^2}=0~.
	\label{ode}
\end{equation}
It will be convenient to write the extremality condition as
\begin{align}
	\p_x \ln f-x_c'\p_y \ln f-\frac{2x_c''}{1+x_c'^2}&=0~,&\partial_{i} \ln f &=3\frac{\partial_{i} h_1}{h_1}+3\frac{\partial_{i} h_2}{h_2}+\frac{\partial_{i} W}{W}~.
\end{align}
We note that $f$ generally vanishes on $\partial\Sigma$ for regular solutions, so $\partial_y \ln f$ diverges. With the Neumann b.c.\ in (\ref{eq:Neumann-bc}), however, all terms are finite.

\subsection{Interface geometries}\label{sec:interface-geometries}

Depending on the choice of ${\rm AdS}_4$ geometry in the 10d metric (\ref{eq:metric-10d}), the solutions describe different interface geometries. In particular, 
\begin{itemize}
	\item[(i)] planar $\RR^{1,2}$ interfaces correspond to Poincar\'e ${\rm AdS}_4$ with $ds^2_{AdS_4}=z^{-2}(dz^2+ds^2_{\RR^{1,2}})$
	\item[(ii)] $\RR\times S^2$ interfaces correspond to global ${\rm AdS}_4$ with $ds^2_{AdS_4}=d\rho^2-\cosh^2\!\rho\, dt^2+\sinh^2\!\rho\,ds^2_{S^2}$
	\item[(iii)] Euclidean $S^3$ interfaces correspond to Euclidean ${\rm AdS}_4$ with $ds^2_{AdS_4}=d\rho^2+\sinh^2\!\rho\,ds^2_{S^3}$
\end{itemize}
For the discussion of entanglement entropies we focus on the first two cases.
Case (i) corresponds to fig.~\ref{fig:iCFT-planar}, case (ii) to fig.~\ref{fig:iCFT-cylinder} (see also fig.~\ref{fig:geometry} below).
As far as the existence of RT surfaces for half space EE's and their shape in the internal space are concerned, the two cases are equivalent, since the area functional depends on the AdS$_4$ geometry only through an overall factor.

The divergences of the EE -- and correspondingly of the RT surface areas -- are entirely contained in the factor $\Vol_{{\rm AdS}_4, t={\rm const}}$ in (\ref{area-0}). 
We choose cut-offs $z\geq \epsilon$ for Poincar\'e AdS and $\rho\leq \ln\frac{2R}{\epsilon}$ for global AdS, so that the induced metric on the cut-off surface is a cylinder of radius $R$.
The regularized volume for Poincar\'e and global AdS is 
\begin{align}
	\Vol_{{\rm AdS}^{\text{Poincar\'e}}_{4,t={\rm const}}}&=\Vol_{\RR^2}\int_\epsilon^\infty \frac{dz}{z^3}=\frac{1}{2\epsilon^2}\Vol_{\RR^2}
	~,
	\nonumber\\
	\Vol_{{\rm AdS}^{\text{global}}_{4,t={\rm const}}}&=\Vol_{S^2}\int_0^{\ln\frac{2R}{\epsilon}} d\rho\,\sinh^2\!\rho=\frac{2\pi R^2}{\epsilon^2}+2\pi \ln\frac{\epsilon}{2R}+\mathcal O(\epsilon)~,
\end{align}
where $\Vol_{S^2}=4\pi$ in the second line.
For Poincar\'e AdS$_4$, i.e.\ planar interfaces, the only non-trivial term is the $1/\epsilon^2$ divergence, whose coefficient depends on the choice of regulator (in $d>2$ the half space EE is not IR divergent).
The natural choice for extracting regularization scheme independent quantities is global AdS, corresponding to an $\RR\times S^2$ interface, where the coefficient of the logarithmic divergence is scheme independent.
This also provides a geometric IR regulator.

Option (iii), describing spherical interfaces, would be the natural choice for computing the interface contribution to the free energy. This contribution can be obtained from the EE of a spherical region centered on the interface. For BCFTs and the boundary contribution to the free energy this was discussed in \cite{Raamsdonk:2020tin}. The interface free energies can be obtained analogously, but we will not consider that here. Instead, we focus on surfaces computing EE's between two half spaces.

\section{Case studies}\label{sec:examples}

We now discuss half space EE's for a set of 4d ICFTs, using a combination of analytic and numerical methods.
We start with a ``simple" interface which does not host genuinely lower-dimensional degrees of freedom, and then discuss interfaces hosting 3d degrees of freedom.
We focus on interfaces separating 4d $\mathcal N=4$ SYM theories with identical gauge groups for simplicity; interfaces between 4d $\mathcal N=4$ SYM with different gauge groups can be studied along the same lines.

\subsection{Empty interface}\label{sec:10djanus}

We start with the classic Janus interface, which separates two 4d $\mathcal N=4$ SYM theories on half spaces with the same gauge groups but different gauge couplings (as discussed in \cite[sec.~10.3]{DHoker:2007zhm}; earlier studies include \cite{Bak:2003jk,Clark:2004sb,DHoker:2006qeo}).
The interface does not host additional 3d degrees of freedom aside from the restriction of ambient operators to the interface; in that sense the interface is empty.
There are no 5-branes in the brane construction. The harmonic functions can be taken as
\begin{align}
	h_1&=\alpha\cosh(z-\delta)+\rm{c.c.}
	\nonumber\\
	h_2&=-i\hat\alpha\sinh(z+\delta)+\rm{c.c.}
\end{align}
The solution reduces to ${\rm AdS}_5\times S^5$ for $\delta=0$.
The function $f$ defined in (\ref{area-1}) takes the form
\begin{align}\label{eq:f-Janus}
	f&=
	\alpha ^4 \hat\alpha^4\cosh (2 \delta ) \sin ^4(2 y) (\cosh (2 \delta )+\cosh (2 x))^3~.
\end{align}
It is invariant under the reflection $x\rightarrow -x$ for any $\delta$.
It is also invariant under $y\rightarrow \frac{\pi}{2}-y$. 
An extremal surface splitting $\Sigma$ which we can find in closed form is
\begin{align}\label{eq:Janus-min}
	x_c(y)&=0~.
\end{align}
It is invariant under both $\ZZ_2$ symmetries and satisfies the Neumann condition (\ref{eq:Neumann-bc}). This surface is indeed a minimum of the area functional:  $\partial_x^2 f\vert_{x=0}$ is positive for all $y$, implying that fluctuations of the critical surface increase the area.
The area of the surface, via (\ref{area-0}), (\ref{area-1}), is
\begin{align}
	A_\gamma&=16\pi \alpha ^2 \hat\alpha^2 V_{{\rm AdS}_4,t={\rm const}}V_{S_1^2\times S_2^2} \cosh^3\!\delta\, \sqrt{2\cosh (2 \delta )}~.
\end{align}

The form of the function $f$ in (\ref{eq:f-Janus}), which governs the effective geometry probed by minimal curves (see the discussion in fig.~\ref{fig:D5NS5-eff}), suggests that $x=0$ should be the only minimal surface connecting the two boundary components. A numerical investigation of the extremality condition (\ref{ode}) with $f$ in (\ref{eq:f-Janus}), using the method employed in the next subsections, indeed finds (\ref{eq:Janus-min}) as the only such minimal surface. 
This is consistent with having no additional freedom in splitting the interface, in line with the fact that this interface between two ambient CFTs with different couplings does not host genuinely new 3d degrees of freedom.\footnote{This is similar to the 2d Janus solutions investigated in \cite{Karch:2022vot}, which connect multiple asymptotic ${\rm AdS}_3$ regions. They are dual to junctions of 2d CFTs on half spaces which differ in marginal couplings. There are no interface degrees of freedom and only one extremal surface.}

\subsection{D5/NS5 interface}\label{sec:D5NS5}

\begin{figure}
		\subfigure[][]{\label{fig:D5NS5-sol}
		\begin{tikzpicture}[scale=1]
			\shade [right color=3dcolor!100,left color=3dcolor!100] (-0.3,0)  rectangle (0.3,-2);
			
			\shade [ left color=3dcolor! 100, right color=4dcolor! 100] (0.3-0.01,0)  rectangle (2,-2);
			\shade [ right color=3dcolor! 100, left color=4dcolor! 100] (-0.3+0.01,0)  rectangle (-2,-2);

			\draw[thick] (-2,0) -- (2,0);
			\draw[thick] (-2,-2) -- (2,-2);
			\draw[dashed] (2,-2) -- +(0,2);
			\draw[dashed] (-2,-2) -- +(0,2);
			
			\node at (-0.5,-0.5) {$\Sigma$};
			\node at (2.5,-0.65) {\footnotesize $AdS_5$};
			\node at (2.5,-1) {\footnotesize $\times$};
			\node at (2.5,-1.35) {\footnotesize $S^5$};
			
			\node at (-2.5,-0.65) {\footnotesize $AdS_5$};
			\node at (-2.5,-1) {\footnotesize $\times$};
			\node at (-2.5,-1.35) {\footnotesize $S^5$};

			\draw[very thick] (0,-0.08) -- (0,0.08) node [anchor=south] {\footnotesize NS5};
			\draw[thick] (0,-1.92) -- (0,-2.08) node [anchor=north] {\footnotesize D5};
		\end{tikzpicture}
	}\hskip 20mm
	\subfigure[][]{\label{fig:D5NS5-brane}
	\begin{tikzpicture}[y={(0cm,1cm)}, x={(0.707cm,0.707cm)}, z={(1cm,0cm)}, scale=1.1]
		\draw[gray,fill=gray!100,rotate around={-45:(0,0,1.8)}] (0,0,1.8) ellipse (1.8pt and 3.5pt);
		\draw[gray,fill=gray!100] (0,0,0) ellipse (1.5pt and 3pt);
		
		\foreach \i in {-0.05,0,0.05}{ \draw[thick] (0,-1,\i) -- (0,1,\i);}

		\foreach \i in {-0.075,-0.025,0.025,0.075}{ \draw (-1.1,\i,1.8) -- (1.1,\i,1.8);}
		
		\foreach \i in  {-0.075,-0.045,-0.015,0.015,0.045,0.075}{ \draw (0,1.4*\i,0) -- (0,1.4*\i,1.8+\i);}
		
		\foreach \i in {-0.045,-0.015,0.015,0.045}{ \draw (0,1.4*\i,1.8+\i) -- (0,1.4*\i,4);}
		
		\foreach \i in  {-0.045,-0.015,0.015,0.045}{ \draw (0,1.4*\i,0) -- (0,1.4*\i,-1.8);}

		\node at (-0.18,-0.18,3.4) {\scriptsize $N_{\rm D3}^\infty$ D3};
		\node at (-0.18,-0.18,-0.9) {\scriptsize $N_{\rm D3}^\infty$ D3};
		\node at (1.0,0.3,1.8) {\scriptsize $N_5$ D5};
		\node at (0,-1.25) {\footnotesize $N_5$ NS5};
		\node at (0.18,0.18,0.8) {{\scriptsize $N_{\rm D3}^0$ D3}};
	\end{tikzpicture}
	}
	\caption{Left: Schematic illustration of the supergravity solution (\ref{eq:D5NS5-h12}). The dashed lines correspond to $\Re(z)\rightarrow \pm\infty$. Right: The brane construction with the 5-branes separated for illustration. Within each group of 5-branes all branes have the same net number of D3-branes ending on them.}
\end{figure}
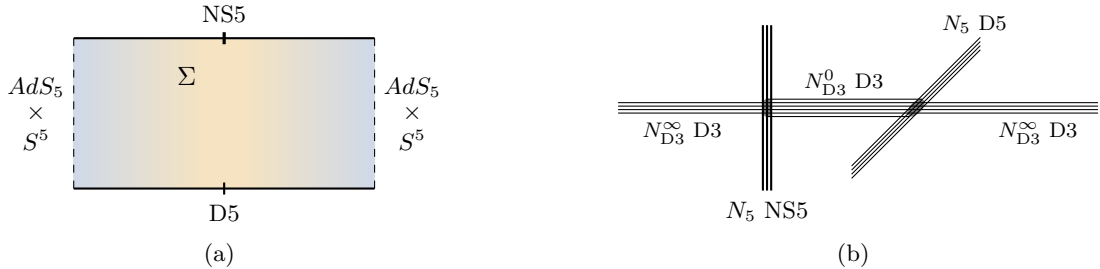

\begin{figure}
	\centering
	\begin{tikzpicture}
		\node at (-8,0) {	\includegraphics[width=0.38\linewidth]{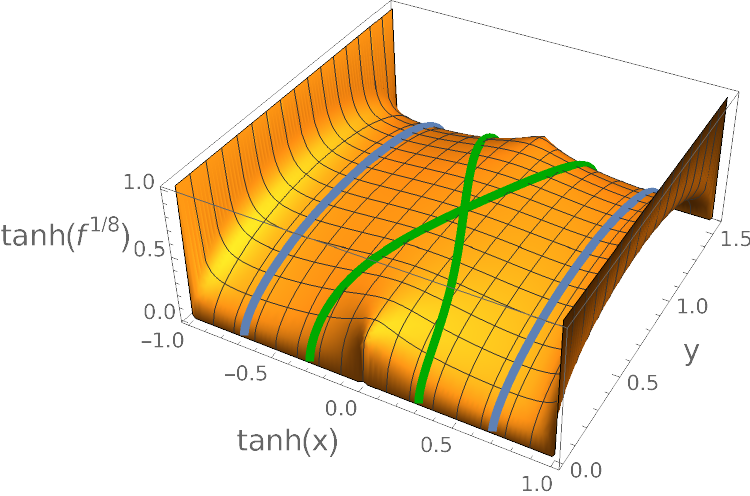}};
		\node at (0,0) {	\includegraphics[width=0.38\linewidth]{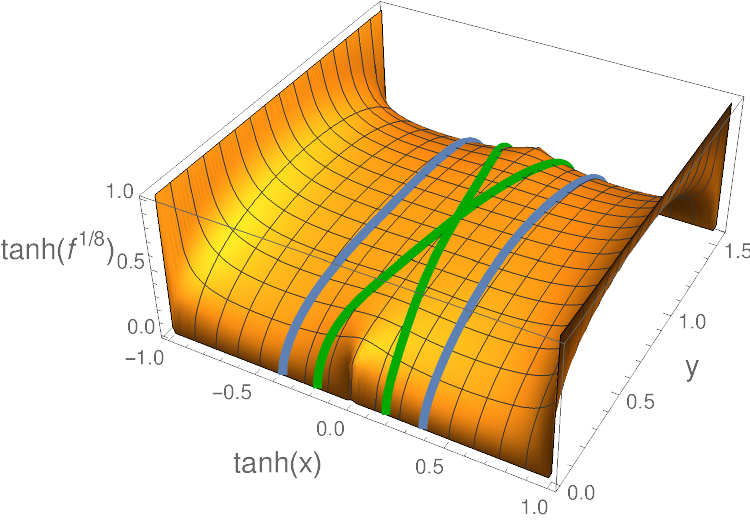}};
	\end{tikzpicture}
	
	\caption{The area functional in (\ref{area-1}) can be understood as arising from an effective 10d geometry with metric $ds^2=dx^2+dy^2+2f(x,y)^{1/8}ds^2_{\mathcal M_8}$, which governs the shape of the minimal surfaces. The plots show the effective warp factor $f^{1/8}$ with the minimal surfaces in blue and green, on the left for $K/N_5=\frac{3}{20}$, on the right for $K/N_5=\frac{11}{20}$.
	At $x\rightarrow\pm \infty$ or $\tanh(x)=\pm1$, the warp factor diverges and the geometry becomes ${\rm AdS}_5\times S^5$. The 5-brane sources are at $x=0$.\label{fig:D5NS5-eff}}
\end{figure}

The next example is an interface engineered by D3-branes intersecting a combination of $N_5$ D5 branes and $N_5$ NS5 branes. This configuration engineers an interface which hosts genuinely 3d degrees of freedom, and this example will serve to illustrate the main qualitative points. We keep the same 4d couplings on both sides of the interface for simplicity, noting that the generalization to different 4d couplings amounts to a minor modification of $h_{1/2}$.
The harmonic functions are
\begin{align}\label{eq:D5NS5-h12}
	h_1&=\frac{\pi\alpha'}{2}K\cosh z-\frac{\alpha'}{4}N_5\ln\tanh \left(\frac{z}{2}\right)+\mathrm{c.c.}
	\nonumber\\
	h_2&=-\frac{i\pi\alpha'}{2}K\sinh z-\frac{\alpha'}{4}N_5\ln\tanh \left(\frac{i\pi}{4}-\frac{z}{2}\right)+\mathrm{c.c.}
\end{align}
The differentials $\partial h_1$ and $\partial h_2$ have poles at $z=0$ and $z=\frac{i\pi}{2}$, respectively, representing the 5-brane sources. The structure of the solution on $\Sigma$ is illustrated in fig.~\ref{fig:D5NS5-sol}.
The harmonic functions and hence the solutions have a reflection symmetry under $x\rightarrow -x$, where $z=x+iy$. 
The reflection $y\rightarrow \frac{\pi}{2}-y$ exchanges $h_1$ and $h_2$ and is a symmetry of the Einstein frame metric (it realizes S-duality).
We will not give the function $f$ in (\ref{area-1}) explicitly, but note that it is positive and finite in the interior of $\Sigma$ and vanishes along the entire boundary of $\Sigma$. Plots can be found in fig.~\ref{fig:D5NS5-eff}. 

The numbers of semi-infinite D3-branes can be extracted from the radii of the asymptotic ${\rm AdS}_5\times S^5$ geometries. The numbers of D3-branes ending on each group of 5-branes can be determined from the supergravity solution following \cite{Assel:2011xz}. We find $N_{\rm D3}^\infty=2N_5K+\pi K^2$ and $N_5/2$ D3-branes ending on each 5-brane. This leads to the brane configuration in fig.~\ref{fig:D5NS5-brane} with
\begin{align}
	N_{\rm D3}^\infty&=2N_5K+\pi K^2~,&
	N_{\rm D3}^0&=\frac{N_5^2}{2}+N_{\rm D3}^\infty~.
\end{align}
The ICFT emerges when collapsing the combination of  D5 and NS5 branes to an intersection at a point, as in fig.~\ref{fig:iCFT-brane}. A UV description as a quiver gauge theory can be obtained by pulling the NS5 branes apart and relocating the D5-branes via Hanany-Witten transitions. This leads to
\begin{align}\label{eq:D5NS5K-quiver-2}
	\widehat{U(N_{\rm D3}^\infty)}- U(N_{\rm D3}^\infty+R)- U(N_{\rm D3}^\infty+2R)-\ldots - U(&N_{\rm D3}^\infty+R^2) - \ldots - U(N_{\rm D3}^\infty+R) - \widehat{U(N_{\rm D3}^\infty)}
	\nonumber\\
	&\ \ \ \ \ \vert	\nonumber\\
	& \ \ \ [N_5]	
\end{align}
where $R=N_5/2$.
The hatted nodes denote 4d $\mathcal N=4$ SYM on a half space; the unhatted nodes denote 3d gauge nodes. 
For a detailed discussion of the coupling between the 3d and 4d gauge nodes we refer to \cite{Gaiotto:2008sa,Gaiotto:2008sd}.

\bigskip
\textbf{Minimal surfaces:}
We now discuss minimal surfaces separating the two half spaces. To accomplish this they have to stretch from one boundary of $\Sigma$ to the other. To find the minimal surfaces we adapt the relaxation method described in \cite[sec.~4]{Uhlemann:2021nhu}. We start it with a trial surface which connects the two boundaries of $\Sigma$ and satisfies the boundary conditions (\ref{eq:Neumann-bc}), and then dynamically evolve the surface to settle onto a minimum of the area functional. The initial surfaces can be chosen as even or odd under vertical reflection on the strip (since the area functional is invariant under this reflection the solutions to the extremality condition have definite parity). One has to choose to which side of the 5-brane source on a given boundary component the trial surface should be anchored, and we run the relaxation for a sample of trial surfaces for each choice.

Due to the $x\rightarrow -x$ reflection symmetry of the 10d solution, $x(y)=0$ is an extremum of the effective area functional (\ref{area-1}). However, since the curve connects the  5-brane sources at $z=0$ and $z=\frac{i\pi}{2}$, the associated surface does not close off: at the end points of the curve the 10d geometry approaches the near-horizon behavior of 5-branes, and the surface extends into the 5-brane regions.
This would call for a reevaluation of the boundary conditions (\ref{eq:Neumann-bc}).
From the perspective of the area functional (\ref{area-1}) with (\ref{eq:Neumann-bc}), $x(y)=0$ is not a minimum, and though fluctuations in AdS can be stable with negative mass-squared, the associated surface is not stable in that sense either (see footnote \ref{foot:stab}).
We focus in the following on surfaces which close off smoothly and are minima of the effective area functional. Since $x=0$ is the only fixed locus of the reflection symmetry, any further extremal surfaces come in pairs: if $x_c(y)$ is an extremal surface, so is $-x_c(y)$, with identical area.

\begin{figure}
	\centering
		\subfigure[][]{\label{fig:D5NS5-plot-1}
				\includegraphics[width=0.31\linewidth]{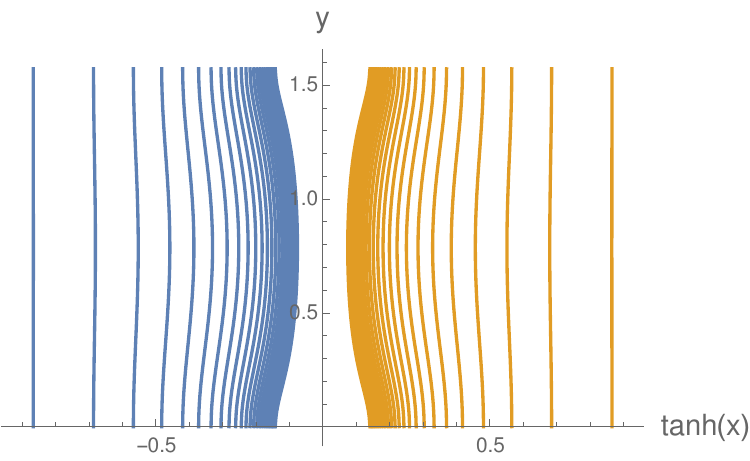}
		}\hskip 0mm
		\subfigure[][]{\label{fig:D5NS5-plot-2}
				\includegraphics[width=0.31\linewidth]{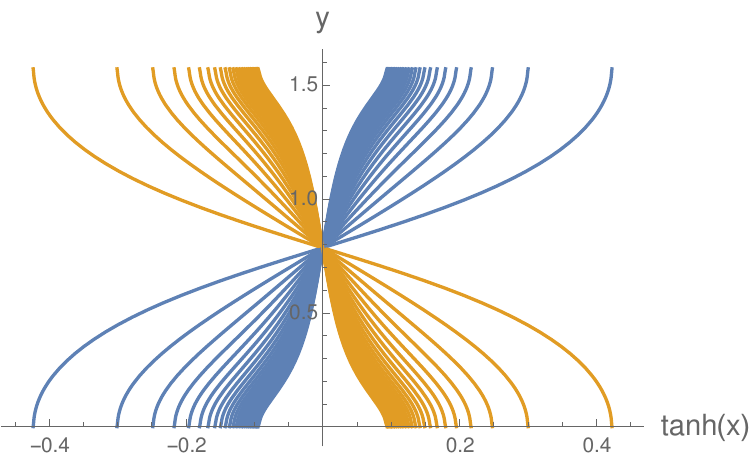}
		}\hskip 0mm
		\subfigure[][]{\label{fig:D5NS5-plot-3}
				\includegraphics[width=0.31\linewidth]{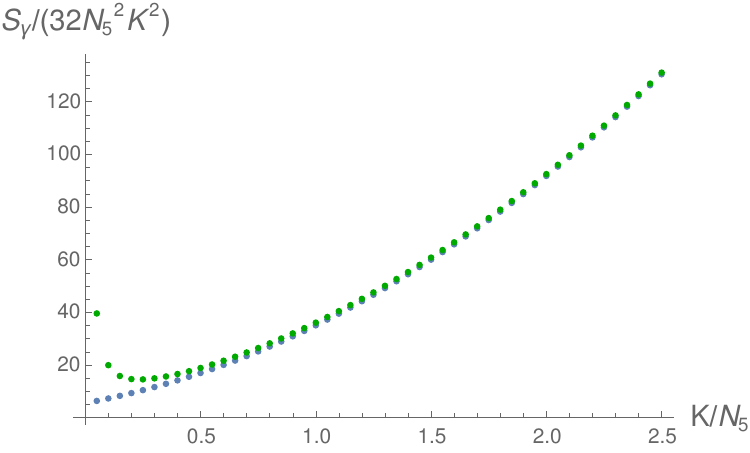}
		}

	\caption{Left: $\ZZ_2^y$ symmetric minimal surfaces $x_c(y)$ from outer to inner for increasing $K/N_5\in\lbrace\frac{1}{20},\ldots,\frac{50}{20}\rbrace$. Center: $\ZZ_2^y$ odd minimal surfaces again from outer to inner for increasing $K/N_5$. Right: The area $S_\gamma$ defined in (\ref{area-0}) divided by $32N_5^2K^2$, in blue for $\ZZ_2^y$ even surfaces and in green for $\ZZ_2^y$ odd surfaces.
	\label{fig:D5NS5-plot}}
\end{figure}

Minimal surfaces found by the relaxation method are shown in fig.~\ref{fig:D5NS5-plot} and fig.~\ref{fig:D5NS5-eff}. 
We find two types of surfaces, which may be grouped according to their behavior under the reflections $x\rightarrow -x$ and $y\rightarrow \frac{\pi}{2}-y$. For each fixed $K/N_5$ there is a pair of surfaces which are symmetric under the $y$ reflection and exchanged under reflection of $x$ (fig.~\ref{fig:D5NS5-plot-1}). These surfaces separate one asymptotic ${\rm AdS}_5\times S^5$ region from the 5-brane region and the second ${\rm AdS}_5\times S^5$ region. The second type of surface is symmetric under simultaneous reflections of $x$ and $y$ (fig.~\ref{fig:D5NS5-plot-2}). These surfaces also come in pairs which are exchanged under $x$ or $y$ reflections.
The latter surfaces separate one asymptotic ${\rm AdS}_5\times S^5$ region together with one 5-brane source from the rest of $\Sigma$.
The area functional (\ref{area-1}) can be understood in the sense of an effective geometry; this is shown and discussed in fig.~\ref{fig:D5NS5-eff}.

For the surfaces symmetric under $y$-reflection (fig.~\ref{fig:D5NS5-plot-1}) the limiting cases $K/N_5 \gg 1$ and $K/N_5\ll 1$ can be understood analytically.
The asymptotic form of the minimal surfaces for $K/N_5\ll 1$ is derived in app.~\ref{app:asympt}. It is given by
\begin{align}
	\pm x_c(y)\big\vert_{K/N_5\ll 1}&=\frac{1}{2}\log\left(\frac{2N_5}{\pi K}\right)+\frac{\pi K}{3N_5}+\mathcal O(K^2/N_5^2)~.
\end{align}
That is, the surface becomes independent of $y$.
In the opposite limit, when $K\gg N_5$, the 5-branes reduce to probes on the ${\rm AdS}_5\times S^5$ geometry and the minimal surfaces both approach $x(y)=0$ (without reaching the limiting surface $x=0$ for finite $K/N_5$).
Both features can be seen in fig.~\ref{fig:D5NS5-plot-1}.

The areas of the surfaces are shown in fig.~\ref{fig:D5NS5-plot-3}. By construction, the surfaces found by the relaxation method are (at least) local minima of the area functional.
The surfaces which are odd under reflection of $y$ generally have larger area than the even surfaces. For large $K/N_5$ the difference vanishes, while for small $K/N_5$ it grows. The area of the odd surfaces actually diverges for $K/N_5\rightarrow 0$. This can be understood as follows: The anchor points on $\partial\Sigma$ are pulled out to $\Re(z)\rightarrow\pm \infty$ for $K/N_5\rightarrow 0$ for both types of surfaces. For the even surfaces in fig.~\ref{fig:D5NS5-plot-1}, both anchor points are pulled in the same direction. For the odd surfaces in in fig.~\ref{fig:D5NS5-plot-2}, on the other hand, the anchor points are pulled in opposite directions, stretching the curves to ever larger length.

\bigskip
\textbf{Interpretation:} 
One may wonder whether the 4 different curves in $\Sigma$ for given $K/N_5$ in fig.~\ref{fig:D5NS5-plot-1}, \ref{fig:D5NS5-plot-2} should be considered as physically distinct, satisfying different asymptotic boundary conditions, or compete with each other. 
Each curve corresponds to a local minimum of the effective area functional $S_\gamma$ in (\ref{area-1}).
It may seem tempting to seek surfaces with globally minimal $S_\gamma$.\footnote{Since the curve $x(y)=0$ is not a global minimum of $S_\gamma$, insisting on having a global minimum of the effective area functional would still leave (at least) a pair of minimal surfaces related by the $x\rightarrow -x$ reflection symmetry.}
However, the actual surfaces are 8-dimensional and extend along AdS$_4$. They are anchored at the conformal boundary of AdS$_4$ so as to split the internal space in a particular way. 
They differ in asymptotic boundary conditions.
The actual areas are divergent, due to the AdS$_4$ volume in (\ref{area-0}). The different surfaces corresponding to different curves in $\Sigma$ differ in the divergent parts of their area. This is similar to the usual geometric EE's associated with different regions in the CFT, which correspond to different decompositions of the Hilbert space.
In summary, the 8d surfaces corresponding to different curves in fig.~\ref{fig:D5NS5-plot} are physically distinct. We propose to interpret them as corresponding to different decompositions of the ICFT degrees of freedom.

The data specifying which boundary segment on $\Sigma$ a minimal surface ends on may then be interpreted in the brane construction.
The surfaces in fig.~\ref{fig:D5NS5-plot-1} would naturally correspond to assigning the 5-branes and the associated 3d degrees of freedom to one half space of the ambient CFT, while the surfaces in fig.~\ref{fig:D5NS5-plot-2} would correspond to decompositions where the D5 and NS5 poles are assigned to different half spaces, and the interface degrees of freedom are divided.

The symmetry of the 10d solution under $x\rightarrow -x$ exchanges the two half spaces hosting the 4d $\mathcal N=4$ SYM degrees of freedom, which emerge at $x\rightarrow\pm\infty$.
The absence of a surface respecting this symmetry suggests that there is no decomposition of the ICFT Hilbert space which is symmetric under exchange of the left and right Hilbert spaces associated with the two ambient half spaces. 
However, the reflection $y\rightarrow \frac{\pi}{2}-y$ amounts to performing S-duality (exchanging D5 and NS5 sources on the two boundary components). The $\ZZ_2$ odd surfaces are invariant under the composition of the $\ZZ_2^x$ and $\ZZ_2^y$ reflections, suggesting a decomposition which is symmetric up to S-duality exists.

\bigskip
\textbf{Relation to ambient EE's:} 
We expect the surfaces in fig.~\ref{fig:D5NS5-plot-1} to arise as limiting cases of geometric EE's in the ambient CFTs. One starts with an interval entirely inside one of the half spaces away from the interface, i.e.\ in one of the colored regions in fig.~\ref{fig:iCFT}.
Such EE's are computed holographically by RT surfaces anchored on the boundaries of the interval. In fig.~\ref{fig:D5NS5-sol} these are surfaces anchored with both ends in the same asymptotic ${\rm AdS}_5\times S^5$ region. That is, either entirely on the right end of $\Sigma$ or entirely on the left end. Upon increasing the interval size, one anchor moves towards the conformal boundary of the AdS$_4$ slices at the end of $\Sigma$ while the other anchor drops into the bulk of the AdS$_4$ slices. The RT surface stretches out.
As the region eventually approaches an entire half space in fig.~\ref{fig:iCFT}, the RT surface should approach one of the outer surfaces in fig.~\ref{fig:D5NS5-plot-1}. In fig.~\ref{fig:iCFT-hol-0-b} this would correspond to the green semi-circle approaching one of the dotted red surfaces.
This limit consideration supports an EE interpretation of the surfaces in fig.~\ref{fig:D5NS5-plot}.

The limiting surfaces obtained from large intervals on different sides of the interface are not expected to agree. From an interval in the left/right half space we expect to get the left/right outer surface in fig.~\ref{fig:D5NS5-plot-1}. The existence of multiple minimal surfaces splitting $\Sigma$ is consistent with the fact that there are 3d d.o.f.\ localized on the interface, corresponding to non-empty $\mathcal H_{3d}$ in (\ref{eq:H-ICFT}). The two outer surfaces in fig.~\ref{fig:D5NS5-plot} are then natural candidates for splits which assign the interface degrees of freedom as much as possible to either the left half space or to the right half space.

\subsection{D5$^2$/NS5$^2$ interface}\label{sec:D52-NS52}

\begin{figure}
		\subfigure[][]{\label{fig:D52NS52-sol}
		\begin{tikzpicture}[scale=1]
			\shade [right color=3dcolor!100,left color=3dcolor!100] (-0.3,0)  rectangle (0.3,-2);
			
			\shade [ left color=3dcolor! 100, right color=4dcolor! 100] (0.3-0.01,0)  rectangle (2,-2);
			\shade [ right color=3dcolor! 100, left color=4dcolor! 100] (-0.3+0.01,0)  rectangle (-2,-2);
				
			\draw[thick] (-2,0) -- (2,0);
			\draw[thick] (-2,-2) -- (2,-2);
			\draw[dashed] (2,-2) -- +(0,2);
			\draw[dashed] (-2,-2) -- +(0,2);
			
			\node at (-0.5,-0.5) {$\Sigma$};
			\node at (2.5,-0.65) {\footnotesize $AdS_5$};
			\node at (2.5,-1) {\footnotesize $\times$};
			\node at (2.5,-1.35) {\footnotesize $S^5$};
			
			\node at (-2.5,-0.65) {\footnotesize $AdS_5$};
			\node at (-2.5,-1) {\footnotesize $\times$};
			\node at (-2.5,-1.35) {\footnotesize $S^5$};

			\draw[very thick] (-0.7,-0.08) -- (-0.7,0.08) node [anchor=south] {\footnotesize NS5};
			\draw[very thick] (0.7,-0.08) -- (0.7,0.08) node [anchor=south] {\footnotesize NS5};
			\draw[thick] (-0.7,-1.92) -- (-0.7,-2.08) node [anchor=north] {\footnotesize D5};
			\draw[thick] (0.7,-1.92) -- (0.7,-2.08) node [anchor=north] {\footnotesize D5};
		\end{tikzpicture}
	}\hskip 15mm
	\subfigure[][]{\label{fig:D52NS52-brane}
	\begin{tikzpicture}[y={(0cm,1cm)}, x={(0.707cm,0.707cm)}, z={(1cm,0cm)}, xscale=1,yscale=1.1]
		\draw[gray,fill=gray!100] (0,0,-0.5) ellipse (1.8pt and 3pt);
		\draw[gray,fill=gray!100] (0,0,1) ellipse (1.8pt and 4.5pt);
		\draw[gray,fill=gray!100,rotate around={-45:(0,0,2.5)}] (0,0,2.5) ellipse (1.8pt and 5pt);
		\draw[gray,fill=gray!100,rotate around={-45:(0,0,4)}] (0,0,4) ellipse (1.8pt and 3pt);				
		
		\foreach \i in {-0.05,0,0.05}{ \draw[thick] (0,-1,-0.5+\i) -- (0,1,-0.5+\i);}
		\foreach \i in {-0.05,0,0.05}{ \draw[thick] (0,-1,1+\i) -- (0,1,1+\i);}

		\foreach \i in {-0.075,-0.025,0.025,0.075}{ \draw (-1.1,\i,2.5) -- (1.1,\i,2.5);}
		\foreach \i in {-0.075,-0.025,0.025,0.075}{ \draw (-1.1,\i,4) -- (1.1,\i,4);}
		
		\foreach \i in {-0.06,-0.03,0,0.03,0.06}{ \draw (0,1.4*\i,-0.5) -- (0,1.4*\i,1);}
		\foreach \i in {-0.1,-0.075,-0.045,-0.015,0.015,0.045,0.075,0.1}{ \draw (0,1.4*\i,1) -- (0,1.4*\i,2.5+\i);}
		\foreach \i in {-0.06,-0.03,0,0.03,0.06}{ \draw (0,1.4*\i,2.5) -- (0,1.4*\i,4);}
		
		\foreach \i in {-0.03,0,0.03}{ \draw (0,1.4*\i,4) -- (0,1.4*\i,5.5);}
		\foreach \i in {-0.03,0,0.03}{ \draw (0,1.4*\i,-0.5) -- (0,1.4*\i,-2);}
		
		\node at (0,-1.25,-0.5) {\scriptsize $N_5/2$ NS5};
		\node at (0,-1.25,1) {\scriptsize $N_5/2$ NS5};
		\node at (1.0,0.35,2.5) {\scriptsize  $N_5/2$ D5};
		\node at (1.0,0.35,4) {\scriptsize  $N_5/2$ D5};
		\node at (0.2,0.2,1.75) {{\scriptsize $N_{\rm D3}^0$}};
		\node at (0,0.28,0.25) {{\scriptsize $N_{\rm D3}^1$}};
		\node at (0,0.28,3.5) {{\scriptsize $N_{\rm D3}^1$}};
		
		\node at (0,0.28,5) {{\scriptsize $N_{\rm D3}^\infty$}};
		\node at (0,0.28,-1.25) {{\scriptsize $N_{\rm D3}^\infty$}};
	\end{tikzpicture}
	}
	\caption{Left: Illustration of the supergravity solutions (\ref{eq:D52NS52-h12}).
		Right: Brane configuration with D3-branes intersecting two groups of D5-branes and two groups of NS5-branes with D3-branes suspended between them. The net numbers of D3-branes ending on each 5-brane differ between the groups for $\delta\neq 0$.}
\end{figure}

The next example is a more general interface where the D5 and NS5 branes each comprise two groups with different numbers of D3-branes ending on them. This leads to an interface with more general 3d degrees of freedom and a new parameter $\delta$. 
We again keep the same 4d coupling on both sides for simplicity, noting that interfaces separating 4d $\mathcal N=4$ SYM theories with different couplings can be realized along the same lines.
The harmonic functions are
\begin{align}\label{eq:D52NS52-h12}
	h_1&=\frac{\pi\alpha'}{2}K\cosh z-\frac{\alpha'}{4}\frac{N_5}{2}\ln\left[\tanh \left(\frac{z-\delta}{2}\right)\tanh \left(\frac{z+\delta}{2}\right)\right]+\mathrm{c.c.}
	\nonumber\\
	h_2&=-\frac{i\pi\alpha'}{2}K\sinh z-\frac{\alpha'}{4}\frac{N_5}{2}\ln\left[\tanh \left(\frac{i\pi}{4}-\frac{z-\delta}{2}\right)\tanh \left(\frac{i\pi}{4}-\frac{z+\delta}{2}\right)\right]+\mathrm{c.c.}
\end{align}
Two pairs of 5-brane sources are at $z=\pm\delta$ and $z=\pm \delta+\frac{i\pi}{2}$.
For $\delta=0$ the solution reduces to (\ref{eq:D5NS5-h12}). The functions $h_{1/2}$ and the 10d solutions are again invariant under $x\rightarrow -x$ and $y\rightarrow \frac{\pi}{2}-y$.

The brane configuration is shown in fig.~\ref{fig:D52NS52-brane}. 
The numbers of D3-branes ending on each 5-brane are identical within each group shown in the figure, but for $\delta\neq 0$ they differ between the two groups of D5 and NS5 branes.
The D3-brane numbers obtained using \cite[(4.27)]{Assel:2011xz} are
\begin{align}
	N_{\rm D3}^\infty&=\pi K^2+2N_5 K \cosh \delta~, &
	N_{\rm D3}^0&=N_{\rm D3}^\infty+\frac{N_5^2}{2}~,& 
	N_{\rm D3}^1&=N_{\rm D3}^\infty+N_5^2\frac{\Delta_{k,\delta}}{4}~, 
\end{align}
where we defined $k\equiv K/N_5$ and
\begin{align}
	\Delta_{k,\delta}&=\frac{1}{2}+\frac{2}{\pi}\arctan e^{-2\delta}-4k\sinh\delta~.
\end{align}
We note that while $\Delta_{k,\delta}$ can be positive or negative, $N_{\rm D3}^1$ is positive.
The duals for the 3d SCFTs obtained for $K=0$ were used in \cite[sec.~6]{Uhlemann:2021nhu} for a string theory realization of wedge holography. Detailed discussions of the 3d SCFTs at $K=0$ can be found in \cite{Coccia:2020wtk,Coccia:2021lpp}. Here we focus on the supergravity solutions dual to ICFTs for non-zero $K$.

The ICFT can again be obtained as IR fixed point of a mixed 3d/4d quiver gauge theory.
The form of the quiver depends on the sign of $\Delta_{k,\delta}$. We focus on $\Delta_{k,\delta}>0$.
Using a hat to denote the 4d gauge nodes and $s=N_5\Delta_{k,\delta} /2$ the quiver then takes the form
\begin{align}
	\widehat{(N_{\rm D3}^\infty)}- (N_{\rm D3}^\infty+s)-\ldots - (&N_{\rm D3}^\infty+s^2) - \ldots - (X) - \ldots - (N_{\rm D3}^\infty+s^2) - \ldots - (N_{\rm D3}^\infty+s) - \widehat{(N_{\rm D3}^\infty)}
	\nonumber \\
	&\quad\ \vert\  \qquad  \qquad \qquad \qquad \quad \quad   \qquad\quad \vert
	\nonumber\\
	 & \left[N_5/2\right]	\qquad \qquad \quad\qquad \qquad \qquad
	\left[N_5/2\right]
\end{align}%
Along the first ellipsis the 3d gauge group ranks increase in steps of $s$. Along the second ellipsis they decrease in steps of $N_5/2-s$. At the central node $X=N_{\rm D3}^\infty+\frac{1}{4}N_5^2(2\Delta_{k,\delta}-1)$. Along the third ellipsis the ranks increase in increments of $N_5/2-s$ and along the fourth they decrease by $s$.

\bigskip
\textbf{Minimal surfaces:}
An exhaustive analysis of the minimal surfaces in this geometry is beyond our intentions here. We will discuss a selection of surfaces and features, to illustrate that with increasing number of 5-brane sources the effective geometry seen by the surfaces becomes more involved and there are generally more options for local minima of the effective area functional. 

\begin{figure}
	\subfigure[][]{\label{fig:delta-c}
	\begin{tikzpicture}
		\node at (0,0) {
			\includegraphics[width=0.33\linewidth]{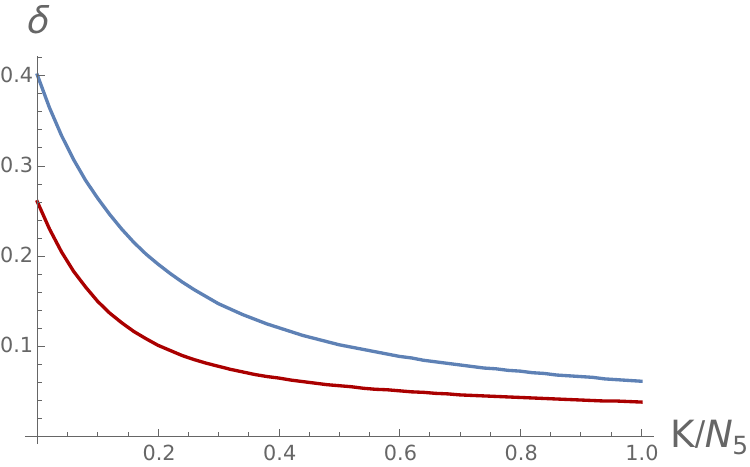}
		};
		\node at (1,-0.8) {\scriptsize $\delta_{\star}$};
		\node at (1,-1.33) {\scriptsize $\delta_{\rm BF}$};
	\end{tikzpicture}
	}\hskip -9mm
	\subfigure[][]{\label{fig:D52NS52-e}
		\includegraphics[width=0.33\linewidth]{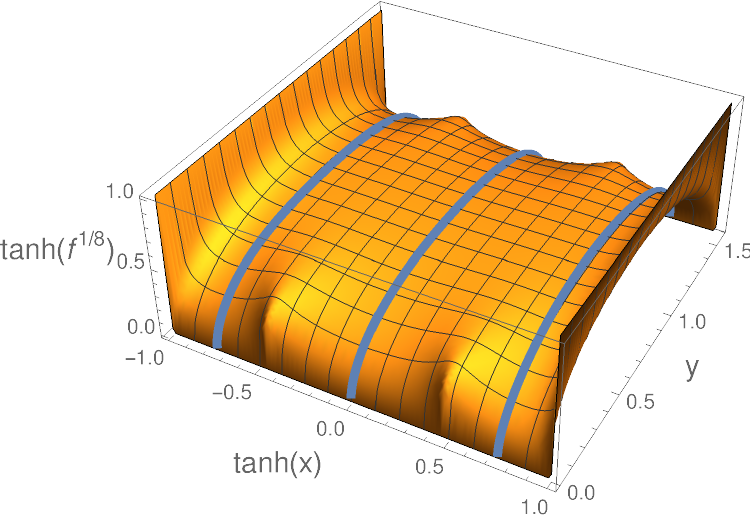}
	}\hskip 1mm
		\subfigure[][]{\label{fig:D52NS52-f}
		\includegraphics[width=0.33\linewidth]{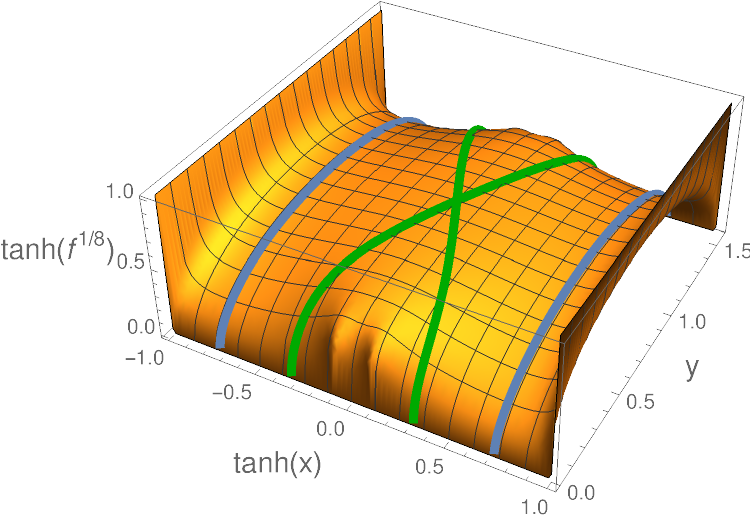}
	}
	\caption{Left: Above/below the upper blue curve $x(y)=0$ is a local minimum/saddle point of the effective area functional $S_\gamma$. Above/below the lower red curve, the lowest-lying fluctuation of the surface on $\rm AdS_3$ has mass-squared above/below the $\rm AdS_3$ BF bound.
	Center: Minimal surfaces for $K/N_5=1/8$ and $\delta=1/2$, where $x=0$ is a minimal surface.
	Right: Minimal surfaces for $K/N_5=1/8$ and $\delta=1/10$.}
\end{figure}

The 10d geometry again has $x\rightarrow -x$ as reflection symmetry and $x(y)=0$ is an extremal surface.
For $\delta>0$ it closes off smoothly at the boundaries of $\Sigma$ and separates the 5-brane sources. Whether it is a minimum of the effective area functional depends on $\delta$. For large enough $\delta$ it is a minimum, while for small $\delta$ some transverse fluctuation modes have negative mass-squared.
The transition value, $\delta_{\star}$, depends on $K/N_5$.  
Since the surfaces wrap a (Euclidean) $\rm AdS_3$ in the $\rm AdS_4$ part of the geometry, we may seek extremal surfaces which are stable in the $\rm AdS_3$ sense and allow for fluctuations with negative mass-squared above the $\rm AdS_3$ Breitenlohner-Freedman bound. This entails a calculation of the fluctuation spectrum, which we outline in app.~\ref{app:fluctuations}. 
The results reveal another transition, at a transition value $\delta_{\rm BF}$, above/below which the lowest-lying fluctuation mode on $\rm AdS_3$ is above/below the BF bound.
For $K/N_5\rightarrow 0$,
\begin{align}\label{eq:delta-crit-K0}
	\delta_{\rm BF}\big\vert_{K/N_5\ll 1}&\approx 0.26~,
	&
	\delta_{\star}\big\vert_{K/N_5\ll 1}&\approx 0.40~.
\end{align}
As $K/N_5$ increases, $\delta_{\rm BF}$ and $\delta_{\star}$ decrease, as shown in fig.~\ref{fig:delta-c}.\footnote{For $\delta\rightarrow0$, when the setup approaches the one considered in sec.~\ref{sec:D5NS5}, the $x=0$ surface is not stable in either sense.\label{foot:stab}}
For $N_5\rightarrow 0$ the solution becomes $\rm AdS_5\times S^5$ and the $x=0$ surface is stable.
We will not discuss the fate of unstable surfaces\footnote{For $K=0$ the surfaces can bend and close off at the ends of the strip, which can lead to the island surfaces of \cite[sec.~6]{Uhlemann:2021nhu}. For $K>0$ the surfaces can not close off a the ends of the strip.} and focus on minimal surfaces in the strictest sense, i.e.\ local minima of the effective area functional (\ref{area-1}),  noting that there may be additional minimal surfaces.
As discussed in sec.~\ref{sec:D5NS5}, RT surfaces associated with different local minima of $S_\gamma$ compute different field theory quantities.

Minimal surfaces other than $x=0$ again come in pairs $\pm x_c(y)$ related by the reflection symmetry. 
As before, we can find surfaces cutting off the asymptotic regions of $\Sigma$ outside the region which contains the 5-brane sources. 
An example are the outer surfaces in fig.~\ref{fig:D52NS52-e}.
For $k\equiv K/N_5\ll 1$ and fixed $\delta$ they take the form (app.~\ref{app:asympt})
\begin{align}\label{eq:D52NS52-xc}
\pm x_c(y)\big\vert_{K/N_5\ll 1}&=\frac{1}{2}\log\left(\frac{2N_5\cosh\delta}{K\pi}\right)+\frac{\pi K\cosh\delta}{3N_5}+\mathcal O(K^2/N_5^2)~.
\end{align}
For general $K/N_5$ the surfaces depend on $y$. 
These surfaces do not divide the 5-brane sources and are natural candidates for entropies in which all 3d degrees of freedom are assigned to one half space. 
For $K=0$ the asymptotic $\rm AdS_5\times S^5$ regions close off and the 10d solutions are dual to genuine 3d SCFTs; the outer critical surfaces then disappear. 
If we fix $K/N_5$, whether the outer surfaces in fig.~\ref{fig:D52NS52-e} exist as local minima of the effective area functional (\ref{area-1}) depends on $\delta$. They do as long as $\delta$ is not too large.\footnote{Intuitively speaking, the bulges corresponding to the 5-brane sources in fig.~\ref{fig:D52NS52-e} merge into the ${\rm AdS}_5\times S^5$ divergences at the ends of the strip for sufficiently large $\delta$, allowing the outer surfaces to slide to $x=0$.} 
Combined with the $x=0$ surface which exists for large enough $\delta$, we have three minimal surfaces which are symmetric under $y\rightarrow \frac{\pi}{2}-y$, compared to two such surfaces in sec.~\ref{sec:D5NS5}. 
For small $\delta$ we also find $\ZZ_2$-odd surfaces connecting the outer valleys in the effective geometry, as shown in fig.~\ref{fig:D52NS52-f}. They separate the D5-brane sources from the NS5-brane sources and generalize the green surfaces in fig.~\ref{fig:D5NS5-eff}, to which they reduce for $\delta=0$.

Our discussion of supergravity solutions with symmetric arrangements of D5 and NS5 brane sources, which are invariant under S-duality, suggests that solutions with more minimal surfaces for half-space EE's can be realized by  adding further pairs of 5-brane sources, generalizing fig.~\ref{fig:D52NS52-e}. Solutions with less symmetric 5-brane configurations can be discussed along similar lines.

\bigskip

It would be interesting to characterize the interfaces in effective field theory terms. From the microscopic perspective we have the parameters $N_5$, $K$ and $\delta$. One combination fixes the asymptotic ${\rm AdS}_5\times S^5$ radius, i.e.\ the 4d central charge (the asymptotic dilaton controlling the 4d gauge couplings is set to one for simplicity, see \cite[sec.~4]{Karch:2022rvr} for a prescription to generalize the solutions). The remaining two parameter control the interface central charge and the nature of the interface degrees of freedom. In more effective field theory terms this should have a description in terms of transmission and absorption coefficients. It would be interesting to compare to \cite{Bachas:2020yxv,Bachas:2022etu}.

For $K=0$, the semi-infinite D3-branes in the brane setup in fig.~\ref{fig:D52NS52-brane} disappear and the field theory becomes a genuine 3d SCFT. The $K=0$, $\delta>0$ solutions describe 3d SCFTs which naturally decompose into two subsectors. This was the reason for picking this setup in \cite[sec.~6]{Uhlemann:2021nhu} for a string theory realization of wedge holography.  The critical value $\delta_{\rm BF}$ in (\ref{eq:delta-crit-K0}) has an interpretation in the context of the discussion in \cite[sec.~6]{Uhlemann:2021nhu}: it marks a transition where the HM surface disappears as a minimal surface at zero temperature (the $x\neq 0$ minimal surfaces only exist for $K>0$). This is similar to the transition for non-gravitating bath in \cite[sec.~5]{Uhlemann:2021nhu}.

\section{Relations for simple interfaces}\label{sec:universal}

In this section we discuss simple interfaces, meaning interfaces whose holographic duals have only one minimal surface anchored on the interface. In the examples of sec.~\ref{sec:examples} this was true for the `empty' interface which does not host 3d degrees of freedom aside from the restriction of bulk operators to the interface.

With this assumption for the holographic duals, relations between entanglement entropies in 2d ICFTs were derived in \cite{Karch:2021qhd,Karch:2022vot}. In the following we first review the argument at the  conceptual level, and then extend it to 4d ICFTs. We stress that the assumption of a single minimal surface splitting the interface is a restriction; examples which violate it are in sec.~\ref{sec:D5NS5}, \ref{sec:D52-NS52}.

\begin{figure}
	\centering
	\subfigure[][]{\label{fig:geometry-0}
		\begin{tikzpicture}
			\node at (0,0) {\includegraphics[width=0.22\linewidth]{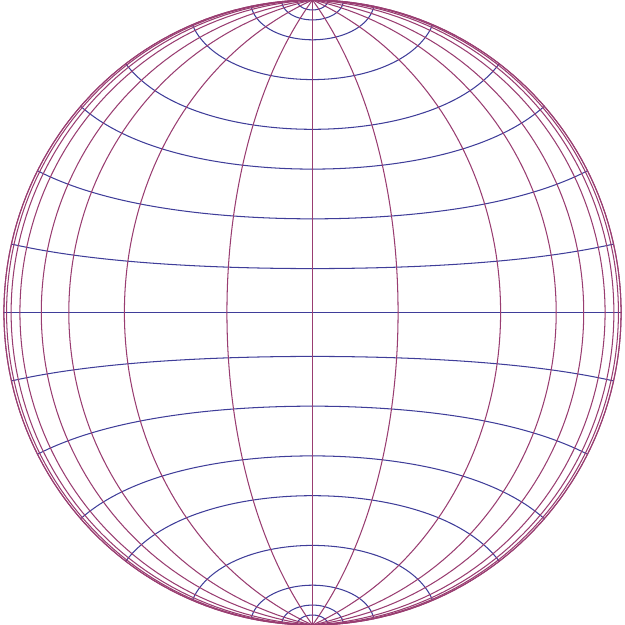}};
			\node at (0,2.0) {\footnotesize $\rho=\infty$};
			\node at (0,-2) {\footnotesize $\rho=\infty$};
			\draw[thick,->] (0.5,0) -- (1.7,0);
			\node at (1.1,0.2) {\scriptsize {\boldmath{$u\rightarrow +\infty$}}};
			\draw[thick,->] (-0.5,0) -- (-1.7,0);
			\node at (-1.1,0.2) {\scriptsize {\boldmath{$u\rightarrow -\infty$}}};
			\draw [very thick,green] (0,-1.8) -- (0,1.8);
		\end{tikzpicture}
	}
	\hskip 10mm
	\subfigure[][]{\label{fig:geometry-1}
		\begin{tikzpicture}
			\node at (0,0) {\includegraphics[width=0.22\linewidth]{adsads-slicing.pdf}};
			\node at (0,2.0) {\footnotesize $\rho=\infty$};
			\node at (0,-2) {\footnotesize $\rho=\infty$};
			\draw[thick,->] (0.5,0) -- (1.7,0);
			\node at (1.1,0.2) {\scriptsize {\boldmath{$u\rightarrow +\infty$}}};
			\draw[thick,->] (-0.5,0) -- (-1.7,0);
			\node at (-1.1,0.2) {\scriptsize {\boldmath{$u\rightarrow -\infty$}}};
			\draw [very thick,green] (0-0.7,1.8-0.12) arc (210:330:0.8);
			\draw [very thick,green] (0+0.7,-1.8+0.12) arc (30:150:0.8);
		\end{tikzpicture}
	}
	\hskip 10mm
	\subfigure[][]{\label{fig:geometry-2}
		\begin{tikzpicture}
			\node at (0,0) {\includegraphics[width=0.22\linewidth]{adsads-slicing.pdf}};
			\node at (0,2.0) {\footnotesize $\rho=\infty$};
			\node at (0,-2) {\footnotesize $\rho=\infty$};
			\draw[thick,->] (0.5,0) -- (1.7,0);
			\node at (1.1,0.2) {\scriptsize {\boldmath{$u\rightarrow +\infty$}}};
			\draw[thick,->] (-0.5,0) -- (-1.7,0);
			\node at (-1.1,0.2) {\scriptsize {\boldmath{$u\rightarrow -\infty$}}};
			\draw [very thick,green] (0,1.8) arc (180:318:0.7);
			\draw [very thick,green] (0,-1.8) arc (180:180-138:0.7);
		\end{tikzpicture}
	}
	\caption{Illustration of the geometry (\ref{eq:5d-metric1}), (\ref{eq:AdS4-global}). Vertical curves correspond to constant $u$, horizontal curves to constant $\rho$. 
		The figures show a section through the geometry with the $S^2$ directions in (\ref{eq:AdS4-global}) suppressed. The $\rho$ coordinate runs through $[0,\infty)$. The upper/lower halves correspond to antipodal points on $S^2$. From left to right the green curves show RT surfaces for case 2, case 1, and case 1b. \label{fig:geometry}}
\end{figure}

The general bottom-up metric for an AdS$_5$/ICFT$_4$ bulk can be written as a warped product of AdS$_4$ over a line with metric
\begin{equation}\label{eq:5d-metric1}
	ds^2=e^{2A(u)}ds_{AdS_4}^2+du^2~.
\end{equation}
The warpfactor $A(u)$ sets the size of the AdS$_4$ slices. 
Two asymptotic AdS$_5$ regions emerge at $u\rightarrow \pm\infty$ where $A\vert_{|u|\gg 1}\sim |u|/L$, with $L$ setting the asymptotic AdS$_5$ radius.
For the 2d ICFT models in \cite{Karch:2021qhd,Karch:2022vot}, Poincar\'e AdS was used on the AdS slices, corresponding to planar interfaces. In 4d, global AdS is the natural choice for extracting scheme independent terms, as discussed in sec.~\ref{sec:interface-geometries}. We therefore use
\begin{equation}\label{eq:AdS4-global}
	ds_{AdS_4}^2=d\rho^2-\cosh^2\!\rho\, dt^2+\sinh^2\!\rho\,ds^2_{S^2}~.
\end{equation}
The ICFT$_4$ geometry is $\RR\times S^3$. The two half ambient CFTs are at $u\rightarrow \pm\infty$ (fig.~\ref{fig:geometry}).
The spatial ICFT geometry comprises two hemispheres $HS^3_L\cup HS^3_R$, where $HS^3_L$ emerges at $u=-\infty$ and is covered by the $(\rho,S^2)$ coordinates, while $HS^3_R$ emerges at $u=+\infty$ and is likewise covered by $(\rho,S^2)$ coordinates.
The 3d interface is located at $\rho\rightarrow\infty$ and has geometry $\RR\times S^2$.

The 2d results of \cite{Karch:2021qhd,Karch:2022vot} relate the divergences of the EE's of two regions, referred to as case 1b and case 2 there. We use a similar notation for 4d ICFTs and define the regions as follows:
\begin{itemize}
	\item Case 1: Region containing the interface defined as $HS^3_L\big\vert_{\rho>\rho_L}\cup HS^3_R\big\vert_{\rho>\rho_R}$ (fig.~\ref{fig:geometry-1}). Case~1b refers to the limiting case $\rho_L\rightarrow\infty$ with $\rho_R$ fixed. This leads to a a one-sided region (fig.~\ref{fig:geometry-2}).
	
	\item Case 2: A half space, say $HS^3_L$. This leads to an EE between CFT$_L$ and CFT$_R$ (fig.~\ref{fig:geometry-0}). 
\end{itemize}
The EE's are computed holographically by the areas of 3d RT surfaces $R_I$ in the bulk, which are anchored on the 2d boundaries of $I$,
\begin{equation}
	S_{\rm EE}=\frac{\mathcal{A}(R_I)}{4G_N}~,
\end{equation}
where $G_N$ is the 5d Newton constant. The surfaces are shown schematically in green in fig.~\ref{fig:geometry}.

The relations between the divergences in the case 1b and case 2 EE's can be understood as follows. Holographically, the divergences originate from the region where the minimal surfaces end on the conformal boundary.
The minimal surface for the half space EE (fig.~\ref{fig:geometry-0}) is anchored only on the interface, and picks up the associated divergences. The minimal surface for the one-sided region (fig.~\ref{fig:geometry-2}) is anchored at two places on the conformal boundary. One is the interface and identical to the anchor of the case 2 surface. The other is in the ambient CFT away from the interface. The additional anchor picks up divergences dictated by the ambient CFT central charge.

The regularization scheme independent quantities in 4d EE's are the coefficients of the logarithmically divergent terms.
With a UV cut-off $\epsilon$ we thus define the coefficients $\sigma_{1/2}$ as
\begin{align}\label{eq:sigma-def}
	S_{\rm EE}^{(1b)}&=\Vol_{S^2}\left[\frac{a_1}{\epsilon^2}+\sigma_1\ln\epsilon+\mathcal O(1)\right]~,
	&
	S_{\rm EE}^{(2)}&=\Vol_{S^2}\left[\frac{a_2}{\epsilon^2}+\frac{\sigma_2}{2}\ln\epsilon+\mathcal O(1)\right]~.
\end{align}
The factor 2 in the definition of $\sigma_2$ follows common conventions in 2d.
We also note that $\Vol_{S^2}=4\pi$ and $\sigma_{1/2}$ are dimensionless.
For interfaces with only one minimal surface splitting the interface in the holographic dual, as described above, we then find (app.~\ref{app:universal})
\begin{equation}\label{eq:simple-rel}
	\sigma_1=\frac{\sigma_2}{2}+\frac{L^3}{8G_N}~.
\end{equation}
The combination of 5d Newton constant and the asymptotic AdS$_5$ curvature radius $L$ is related to the 4d ambient CFT central charge, e.g.\ $\pi L^3/(2G_N)$ is the coefficient of the log term in the EE of a spherical region \cite{Ryu:2006ef}, which is in turn related to the 4d trace anomaly coefficients \cite{Solodukhin:2008dh}. 

The relation (\ref{eq:simple-rel}) is derived in app.~\ref{app:universal} for bottom-up models with geometry (\ref{eq:5d-metric1}). The crucial assumption is that the warpfactor $e^{A(u)}$ has a unique minimum, leading to single solutions for the case 2 and case 1b surfaces. If there were multiple minima, multiple minimal surfaces may exist. From the bottom-up perspective these would all have the same anchor point on the interface; we have seen in the 10d solutions that they should be thought of as anchored in different ways in the internal space (which the bottom-up models do not see).
With the single minimal surface assumption the argument can be extended to top-down 10d solutions, as discussed for 2d ICFTs in \cite{Karch:2022vot}. However, as seen in sec.~\ref{sec:examples} the 10d solutions can have multiple minimal surfaces.

We highlight an assumption in the derivation of (\ref{eq:simple-rel}) for 4d ICFTs which differs from 2d: For large $u$, we assume $e^{2A} \sim \frac{1}{4}L^2e^{2u/L}+L^2/2+\ldots$\,. The first term realizes an asymptotically AdS$_5$ geometry (the coefficient is a matter of convention). We further assume that the first subleading term is fixed by the near-boundary bulk equations of motion to agree with the expansion for AdS$_5$ with $e^{2A}=L^2\cosh^2(u/L)$. In field theory terms this is justified if no relevant operators are turned on in the ambient CFT. In 2d the logarithmic term in the EE is the leading divergence and does not depend on the subleading mode of the bulk metric.
One may generalize  (\ref{eq:5d-metric1}), though, to allow for ambient operators which affect the logarithmic term in the EE to be turned on.

\subsection{General picture}\label{sec:gen}

We derived the relation (\ref{eq:simple-rel}) for EE's in 4d ICFTs which are characterized by having a single minimal surface splitting the interface in the holographic dual. Such ``simple interfaces" are qualitatively consistent with the 2d discussions in \cite{Karch:2021qhd,Karch:2022vot}.
However, we have seen that the single minimal surface assumption is not in general true for the top-down models considered here. Explicit examples are in sec.~\ref{sec:D5NS5}, \ref{sec:D52-NS52}. In these examples the interface hosts genuinely lower-dimensional degrees of freedom and there are multiple minimal surfaces corresponding to half space EE's. 

How then do we expect the existence of multiple minimal surfaces splitting the interface to change the universal relations? Case 1b arises from regions which fully include the interface as limiting case where one boundary of the region retracts to the interface. We expect the resulting case 1b RT surface to be anchored on one of the minimal surfaces splitting the interface. 
If there are multiple minimal surfaces splitting the interface (the interface hosts 3d degrees of freedom), the case 2 half space EE can be defined in multiple ways, depending on which minimal surface is chosen. The anchor of the case 2 RT surface on the interface then does not need to agree with the anchor of the case 1b surface. We therefore expect a more general relation of the form
\begin{equation}\label{eq:interface-spectrum}
	\sigma_1=\frac{\sigma_2}{2}+\frac{L^3}{8G_N}+d~,
\end{equation}
where $d$ in our examples can take values in a discrete set. This ``spectrum" of allowed values for $d$ and their degeneracies then characterizes the interface. 

If there is only one minimal surface splitting the interface, the case 2 RT surface is anchored in the same way on the interface as the case 1b surface. In that case there is only one half space EE and this characterizes what we called ``simple interfaces" with $d=0$. An example was discussed in sec.~\ref{sec:10djanus}.
For the interface in sec.~\ref{sec:D5NS5}, as an example of an interface which hosts genuinely lower-dimensional degrees of freedom, we identified two pairs of minimal surfaces for case 2 (shown in green and blue in fig.~\ref{fig:D5NS5-eff}) with degenerate areas. Each pair leads to the same value of the coefficient $\sigma_2$ in the half space EE's and to the same $d$ in the relation (\ref{eq:interface-spectrum}). We find two allowed values for $d$ which can each be realized in two different ways.

\let\oldaddcontentsline\addcontentsline
\renewcommand{\addcontentsline}[3]{}
\begin{acknowledgments}
	We thank Andreas Karch for valuable discussions. The work of MW was supported, in part, by the U.S.~Department of Energy under Grant DE-SC0022021 and by a grant from the Simons Foundation (Grant 651678, AK).
\end{acknowledgments}
\let\addcontentsline\oldaddcontentsline

\appendix
\renewcommand\theequation{\thesection.\arabic{equation}}

\section{Small $K/N_5$ asymptotics}\label{app:asympt}

We consider the limit $K/N_5\rightarrow 0$ for the solutions of sec.~\ref{sec:D5NS5} ($\delta=0$) and sec.~\ref{sec:D52-NS52}, where the number of semi-indefinite D3-branes is small compared to the D3-branes stretching between the 5-branes. 
We derive the non-zero solution $x_c(y)$. Denote $k\equiv K/N_5\rightarrow 0$. The EOM then depends only on $k$ and $\delta$; it is independent of $N_5$ and $\alpha'$.

The shape of $f(x,y)$ in the $x$-direction is a double well. For every $y$, there is a minimum $x_{\rm min}(y)>0$ that satisfies $\p_xf(x_{min}(y),y)=0$. For $k\ll 1$, this minimum becomes independent of $y$. 
We expect the minimal curve to locate near one of the minima of $f$ regarded as function of $x$. 
Based on this observation we are led to the following candidate for the minimal surface
\begin{align}
	x_c(y)\big\vert_{k\ll 1}&=x_0+\mathcal O(k)~, & 
	x_0&=\frac{1}{2}\log \frac{2\cosh \delta}{k\pi}~.
\end{align}
We will in the following show that this is indeed a solution and obtain some correction terms.

First, notice that $f(x_0,y)$ uniformly approaches 0 as $k\rightarrow 0$. Expanding $f$ near $x_0$ for small $k$ leads to
\begin{align}
	f(x_0+X,y)=\,&\frac{\pi^2}{\sqrt{2}}\sin^2(2y)\cosh\delta \cosh^2X\Bigg[k^2\cosh\delta\cosh X
	\nonumber\\
	&+\frac{\pi k^3}{4}e^{-2X}(4\cosh X+(3-2e^{-2X}\cosh 2\delta)\sinh X+\sinh 3X)+O(k^4)\Bigg]
	\label{fexpand}
\end{align}
Then we expand $\p_xf/f$ and $\p_yf/f$,
\begin{align}
\frac{\p_xf}{f}(x_0+X,y)&=3\tanh X+\frac{e^{-2X}(1 + e^{2 X} - e^{4 X} - 3 e^{6 X} - 
	2 (1 + e^{2X} - e^{4 X}) \cosh 2\delta)\pi k}{4\cosh\delta\cosh^2 X}\,,
\label{fxf}
\\
\frac{\p_yf}{f}(x_0+X,y)&=4\cot 2y+O(k^2)~.
\label{fyf}
\end{align}
Using these expressions, we can expand 
\begin{equation}
x_c(y)=x_0+X(y)=x_0+X_0(y)+X_1(y)k+X_2(y)k^2+\dots
\label{ansatz}
\end{equation} 
and determine $X_i$ order by order from \eqref{ode}. The boundary conditions are $X_i'(0)=X_i'(\pi/4)$. At $O(k^0)$ this leads to
\begin{equation}
3\tanh X_0-4X_0'\cot 2y-\frac{X_0''}{1+X_0'^2}=0.
\end{equation}
The only solution compatible with the boundary conditions is 
\begin{equation}
X_0(y)\equiv 0,
\end{equation}
This shows that the leading-order solution is given by $x_c(y)=x_0$, as claimed.
For $O(k^1)$,
\begin{equation}
3X_1-\pi\cosh\delta-X_1''-4X_1'\cot 2y=0
\end{equation}
The solution satisfying the boundary conditions is 
\begin{equation}
X_1(y)=\frac{\pi\cosh\delta}{3}.
\end{equation}

Summarizing, the minimal surface solution to \eqref{ode}, for small $k$, is
\begin{equation}
x_c(y)=\frac{1}{2}\log\left(\frac{2\cosh\delta}{k\pi}\right)+\frac{\pi k\cosh\delta}{3}+\mathcal O(k^2)~.
\end{equation}

\section{Single surface relation}\label{app:universal}
We derive the relation (\ref{eq:simple-rel}) under the assumption that the warp factor $e^A(u)$ in the metric (\ref{eq:5d-metric1}) has a unique minimum at $u=u^\star$ with minimal value $A^\star$. We consider the EE's for case 1b and case 2, and we focus on the divergent parts. The divergences in the minimal surface areas are determined by near-boundary analysis \cite{Graham:1999pm} and we therefore need to only focus on the contributions from near the end points. The surfaces are shown schematically in fig.~\ref{fig:geometry}. 
We repeat for convenience the full metric with AdS$_4$ slicing in global coordinates
\begin{align}\label{eq:5d-metric-app}
	ds^2&=du^2+e^{2A(u)}ds_{AdS_4}^2~, & 	
	ds_{AdS_4}^2&=d\rho^2-\cosh^2\!\rho\, dt^2+\sinh^2\!\rho\,ds^2_{S^2}~.
\end{align}
This is ${\rm AdS}_5$ for $e^A=L\cosh(u/L)$, but a different geometry otherwise.
We assume that the geometry approaches AdS$_5$ asymptotically for large $u$. That is, 
\begin{align}\label{eq:5d-metric-app-exp}
	e^{2A(u)}\Big\vert_{u\rightarrow \pm\infty}&\sim\frac{1}{4}A_0^2e^{2|u|/A_0} +\frac{1}{2}A_0^2 \ldots~.
\end{align}
The coefficient of the first term can be changed by shifting $u$, which preserves the form of (\ref{eq:5d-metric-app}). The ellipsis denote subleading terms which may depart from the form for AdS$_5$. The term of $\mathcal O(u^0)$ in $e^{2A}$ is assumed to be fixed by the bulk Einstein's equations, following e.g.\ \cite{Skenderis:2002wp}. The underlying assumption is that no bulk matter fields are turned on which would alter the near-boundary expansion at this order. In field theory terms this is justified if no relevant operators are turned on in the ambient CFT.

\bigskip
\textbf{\underline{Case 2:}}
We start with case 2 and parametrize the surface by $u(\rho)$. The anchor point in fig.~\ref{fig:geometry-0} is approached for $\rho\rightarrow\infty$. The surface area is given by
\begin{align}
	\cA&=\Vol_{S^2}\int d\rho \,e^{2A(u)}\sinh^2\!\rho\,\sqrt{e^{2A(u)}+u'^2}~.
\end{align}
From the equation of motion we can immediately find a solution with which is anchored on the boundary only on an equatorial $S^2$ in $S^3$; with $u_\star$ the minimum of $A(u)$ it takes the form
\begin{align}\label{eq:app-case2-u}
	u(\rho)=u_\star~.
\end{align}
Upon introducing a cut-off $\rho<\rho_\epsilon=-\ln (\epsilon/2)$ the area of the surface (\ref{eq:app-case2-u}) becomes
\begin{align}
	S_{\rm EE}^{(2)}&=\frac{\cA}{4G_N}=\frac{\Vol_{S^2}}{8G_N}e^{3A(u_\star)}\left(\frac{1}{\epsilon^2}+\ln\epsilon\right)+\mathcal O(1)~.
\end{align}
The scheme-independent coefficient $\sigma_2$ defined in (\ref{eq:sigma-def}) therefore reads
\begin{align}
	\frac{\sigma_2}{2}&=\frac{e^{3A(u_\star)}}{8G_N}~.
\end{align}

\bigskip
\textbf{\underline{Case 1b:}} The surface for case 1b is illustrated in fig.~\ref{fig:geometry-2}. One part of the divergences of the EE $S_{\rm EE}^{(1b)}$ is determined by the end point on the interface, which coincides with the anchor point of the case 2 surface above. We note that, since the surface in (\ref{eq:app-case2-u}) is by assumption at a minimum of $e^{A(u)}$,  fluctuations have positive mass. This means more general surfaces satisfying the same boundary condition $\lim_{\rho\rightarrow\infty}u(\rho)=u_\star$ can be expanded as $u=u_\star + \mathcal O(e^{-a\rho})$ with $a>2$. As a result, the divergent contribution from the end point on the interval for the case 1b surface is identical to that of the case 2 surface,
\begin{align}\label{eq:app-delta-def}
	\big(S_{\rm EE}^{(1b)}\big)_{\rm div}&=S_{\rm EE}^{(2)}+\delta~,
\end{align}
where $\delta$ is the divergent contribution from the end point away from the interface in fig.~\ref{fig:geometry-2}. The remaining task then is to determine $\delta$.

Near that end point we parametrize the surface by $\rho(u)$. The area functional takes the form
\begin{align}\label{eq:area-1b-app}
	\cA&=\Vol_{S^2}\int du\, e^{2A(u)}\sinh^2\!\rho\,\sqrt{1+e^{2A(u)}\rho'^2}~.
\end{align}
The boundary condition is $\lim_{u\rightarrow\infty}\rho(u)=\rho_0$. 
With the metric in (\ref{eq:5d-metric-app-exp}), from a standard near-boundary analysis we then find
\begin{align}
	 \rho\vert_{u\rightarrow\infty}&\sim\rho_0-2\coth\rho_0\,e^{-2|u|/A_0}+\ldots~.
\end{align}
Feeding this back into the area functional (\ref{eq:area-1b-app}) with a cut-off $u<u_\epsilon=-A_0\ln(\epsilon/2)$ and performing the integration near the anchor point of the surface away from the defect leads to
\begin{align}
	\cA\big\vert_{u\gg 1}&=A_0^3 \Vol_{S^2}\left[\frac{\sinh^3\!\rho_0}{2\epsilon^2}+\frac{1}{2}\ln\epsilon+\ldots\right]~.
\end{align}
The two leading terms fix the divergent contribution $\delta$ in (\ref{eq:app-delta-def}) and the coefficient of the $\ln\epsilon$ term is scheme independent. In summary, we find that the coefficient $\sigma_1$ defined in (\ref{eq:sigma-def}) is given by
\begin{align}
	\sigma_1&=\frac{\sigma_2}{2}+\frac{A_0^3}{8G_N}~.
\end{align}
Upon identifying $A_0$ with the asymptotic AdS$_5$ curvature radius $L$, we arrive at (\ref{eq:simple-rel}).

\section{Fluctuations}\label{app:fluctuations}

In this appendix we discuss the fluctuation spectrum around the $x(y)=0$ extremal surface identified in sec.~\ref{sec:D52-NS52}. We allow the embedding to depend on the spatial $\rm AdS_4$ directions wrapped by the surface. 
The embedding is then specified by $x(y,\vec{v})$, where $\vec{v}$ are the Euclidean $\rm AdS_3$ coordinates. 
Dropping overall constants, the integrand of the area functional (\ref{area-1}) becomes
\begin{align}
	L&=\sqrt{g_{\rm AdS_3}}\sqrt{f(x,y)} \sqrt{1+(\partial_y x)^2+ k(x,y) g_{\rm AdS_3}^{ij}\partial_i x \partial_j x}~,
\end{align}
where $k=4\rho^2/f_4^2$ and $g_{\rm AdS_3}$ is the metric on unit-radius $\rm AdS_3$. Upon expanding around $x=0$, the linearized extremality condition becomes
\begin{align}
	2\partial_y^2 x+2k(0,y)\Delta_{\rm AdS_3} x-
 \frac{\partial_x^2f(0,y)}{f(0,y)}x+\frac{\partial_yf(0,y)}{f(0,y)}\partial_y x&=0~,
\end{align}
with $\Delta_{\rm AdS_3}$ the Laplacian on Euclidean $\rm AdS_3$.
We expand in harmonics on  $\rm AdS_3$ as $x=x_i(y) \phi_i(\vec{v})$ with $\Delta_{\rm AdS_3}\phi_i = m_i^2\phi_i$. The $\rm AdS_3$ BF bound is $m_i^2\geq -1$. Dropping the index $i$ leads to
\begin{align}\label{eq:app-fluct}
	2\partial_y^2 x+2k(0,y)m^2 x-
	\frac{\partial_x^2f(0,y)}{f(0,y)}x+\frac{\partial_yf(0,y)}{f(0,y)}\partial_y x&=0~.
\end{align}
This is a generalized eigenvalue problem which can be solved numerically using pseudo-spectral methods. We used varying resolutions with up to 50 collocation points and verified that the first eigenvalues are very well converged, to produce fig.~\ref{fig:delta-c}.
As a cross check, we verified that the numerical method reproduces the results for $\rm AdS_5\times S^5$, as obtained in sec.~\ref{sec:10djanus} when $\delta=0$, where the fluctuation equation (\ref{eq:app-fluct}) can be solved analytically.

\bibliography{interface}

\begin{thebibliography}{33}%
\makeatletter
\providecommand \@ifxundefined [1]{%
 \@ifx{#1\undefined}
}%
\providecommand \@ifnum [1]{%
 \ifnum #1\expandafter \@firstoftwo
 \else \expandafter \@secondoftwo
 \fi
}%
\providecommand \@ifx [1]{%
 \ifx #1\expandafter \@firstoftwo
 \else \expandafter \@secondoftwo
 \fi
}%
\providecommand \natexlab [1]{#1}%
\providecommand \enquote  [1]{``#1''}%
\providecommand \bibnamefont  [1]{#1}%
\providecommand \bibfnamefont [1]{#1}%
\providecommand \citenamefont [1]{#1}%
\providecommand \href@noop [0]{\@secondoftwo}%
\providecommand \href [0]{\begingroup \@sanitize@url \@href}%
\providecommand \@href[1]{\@@startlink{#1}\@@href}%
\providecommand \@@href[1]{\endgroup#1\@@endlink}%
\providecommand \@sanitize@url [0]{\catcode `\\12\catcode `\$12\catcode
  `\&12\catcode `\#12\catcode `\^12\catcode `\_12\catcode `\%12\relax}%
\providecommand \@@startlink[1]{}%
\providecommand \@@endlink[0]{}%
\providecommand \url  [0]{\begingroup\@sanitize@url \@url }%
\providecommand \@url [1]{\endgroup\@href {#1}{\urlprefix }}%
\providecommand \urlprefix  [0]{URL }%
\providecommand \Eprint [0]{\href }%
\providecommand \doibase [0]{https://doi.org/}%
\providecommand \selectlanguage [0]{\@gobble}%
\providecommand \bibinfo  [0]{\@secondoftwo}%
\providecommand \bibfield  [0]{\@secondoftwo}%
\providecommand \translation [1]{[#1]}%
\providecommand \BibitemOpen [0]{}%
\providecommand \bibitemStop [0]{}%
\providecommand \bibitemNoStop [0]{.\EOS\space}%
\providecommand \EOS [0]{\spacefactor3000\relax}%
\providecommand \BibitemShut  [1]{\csname bibitem#1\endcsname}%
\let\auto@bib@innerbib\@empty
\bibitem [{\citenamefont {Affleck}\ and\ \citenamefont
  {Ludwig}(1991)}]{Affleck:1991tk}%
  \BibitemOpen
  \bibfield  {author} {\bibinfo {author} {\bibfnamefont {I.}~\bibnamefont
  {Affleck}}\ and\ \bibinfo {author} {\bibfnamefont {A.~W.~W.}\ \bibnamefont
  {Ludwig}},\ }\bibfield  {title} {\bibinfo {title} {{Universal noninteger
  'ground state degeneracy' in critical quantum systems}},\ }\href
  {https://doi.org/10.1103/PhysRevLett.67.161} {\bibfield  {journal} {\bibinfo
  {journal} {Phys. Rev. Lett.}\ }\textbf {\bibinfo {volume} {67}},\ \bibinfo
  {pages} {161} (\bibinfo {year} {1991})}\BibitemShut {NoStop}%
\bibitem [{\citenamefont {Friedan}\ and\ \citenamefont
  {Konechny}(2004)}]{Friedan:2003yc}%
  \BibitemOpen
  \bibfield  {author} {\bibinfo {author} {\bibfnamefont {D.}~\bibnamefont
  {Friedan}}\ and\ \bibinfo {author} {\bibfnamefont {A.}~\bibnamefont
  {Konechny}},\ }\bibfield  {title} {\bibinfo {title} {{On the boundary entropy
  of one-dimensional quantum systems at low temperature}},\ }\href
  {https://doi.org/10.1103/PhysRevLett.93.030402} {\bibfield  {journal}
  {\bibinfo  {journal} {Phys. Rev. Lett.}\ }\textbf {\bibinfo {volume} {93}},\
  \bibinfo {pages} {030402} (\bibinfo {year} {2004})},\ \Eprint
  {https://arxiv.org/abs/hep-th/0312197} {arXiv:hep-th/0312197} \BibitemShut
  {NoStop}%
\bibitem [{\citenamefont {Casini}\ \emph {et~al.}(2016)\citenamefont {Casini},
  \citenamefont {Salazar~Landea},\ and\ \citenamefont
  {Torroba}}]{Casini:2016fgb}%
  \BibitemOpen
  \bibfield  {author} {\bibinfo {author} {\bibfnamefont {H.}~\bibnamefont
  {Casini}}, \bibinfo {author} {\bibfnamefont {I.}~\bibnamefont
  {Salazar~Landea}},\ and\ \bibinfo {author} {\bibfnamefont {G.}~\bibnamefont
  {Torroba}},\ }\bibfield  {title} {\bibinfo {title} {{The g-theorem and
  quantum information theory}},\ }\href
  {https://doi.org/10.1007/JHEP10(2016)140} {\bibfield  {journal} {\bibinfo
  {journal} {JHEP}\ }\textbf {\bibinfo {volume} {10}},\ \bibinfo {pages}
  {140}},\ \Eprint {https://arxiv.org/abs/1607.00390} {arXiv:1607.00390
  [hep-th]} \BibitemShut {NoStop}%
\bibitem [{\citenamefont {Jensen}\ and\ \citenamefont
  {O'Bannon}(2016)}]{Jensen:2015swa}%
  \BibitemOpen
  \bibfield  {author} {\bibinfo {author} {\bibfnamefont {K.}~\bibnamefont
  {Jensen}}\ and\ \bibinfo {author} {\bibfnamefont {A.}~\bibnamefont
  {O'Bannon}},\ }\bibfield  {title} {\bibinfo {title} {{Constraint on Defect
  and Boundary Renormalization Group Flows}},\ }\href
  {https://doi.org/10.1103/PhysRevLett.116.091601} {\bibfield  {journal}
  {\bibinfo  {journal} {Phys. Rev. Lett.}\ }\textbf {\bibinfo {volume} {116}},\
  \bibinfo {pages} {091601} (\bibinfo {year} {2016})},\ \Eprint
  {https://arxiv.org/abs/1509.02160} {arXiv:1509.02160 [hep-th]} \BibitemShut
  {NoStop}%
\bibitem [{\citenamefont {Casini}\ \emph {et~al.}(2023)\citenamefont {Casini},
  \citenamefont {Salazar~Landea},\ and\ \citenamefont
  {Torroba}}]{Casini:2023kyj}%
  \BibitemOpen
  \bibfield  {author} {\bibinfo {author} {\bibfnamefont {H.}~\bibnamefont
  {Casini}}, \bibinfo {author} {\bibfnamefont {I.}~\bibnamefont
  {Salazar~Landea}},\ and\ \bibinfo {author} {\bibfnamefont {G.}~\bibnamefont
  {Torroba}},\ }\bibfield  {title} {\bibinfo {title} {{Irreversibility, QNEC,
  and defects}},\ }\href {https://doi.org/10.1007/JHEP07(2023)004} {\bibfield
  {journal} {\bibinfo  {journal} {JHEP}\ }\textbf {\bibinfo {volume} {07}},\
  \bibinfo {pages} {004}},\ \Eprint {https://arxiv.org/abs/2303.16935}
  {arXiv:2303.16935 [hep-th]} \BibitemShut {NoStop}%
\bibitem [{\citenamefont {{Peschel}}(2005)}]{2005JPhA...38.4327P}%
  \BibitemOpen
  \bibfield  {author} {\bibinfo {author} {\bibfnamefont {I.}~\bibnamefont
  {{Peschel}}},\ }\bibfield  {title} {\bibinfo {title} {{Entanglement entropy
  with interface defects}},\ }\href
  {https://doi.org/10.1088/0305-4470/38/20/002} {\bibfield  {journal} {\bibinfo
   {journal} {Journal of Physics A Mathematical General}\ }\textbf {\bibinfo
  {volume} {38}},\ \bibinfo {pages} {4327} (\bibinfo {year} {2005})},\ \Eprint
  {https://arxiv.org/abs/cond-mat/0502034} {arXiv:cond-mat/0502034
  [cond-mat.stat-mech]} \BibitemShut {NoStop}%
\bibitem [{\citenamefont {Sakai}\ and\ \citenamefont
  {Satoh}(2008)}]{Sakai:2008tt}%
  \BibitemOpen
  \bibfield  {author} {\bibinfo {author} {\bibfnamefont {K.}~\bibnamefont
  {Sakai}}\ and\ \bibinfo {author} {\bibfnamefont {Y.}~\bibnamefont {Satoh}},\
  }\bibfield  {title} {\bibinfo {title} {{Entanglement through conformal
  interfaces}},\ }\href {https://doi.org/10.1088/1126-6708/2008/12/001}
  {\bibfield  {journal} {\bibinfo  {journal} {JHEP}\ }\textbf {\bibinfo
  {volume} {12}},\ \bibinfo {pages} {001}},\ \Eprint
  {https://arxiv.org/abs/0809.4548} {arXiv:0809.4548 [hep-th]} \BibitemShut
  {NoStop}%
\bibitem [{\citenamefont {Karch}\ \emph {et~al.}(2021)\citenamefont {Karch},
  \citenamefont {Luo},\ and\ \citenamefont {Sun}}]{Karch:2021qhd}%
  \BibitemOpen
  \bibfield  {author} {\bibinfo {author} {\bibfnamefont {A.}~\bibnamefont
  {Karch}}, \bibinfo {author} {\bibfnamefont {Z.-X.}\ \bibnamefont {Luo}},\
  and\ \bibinfo {author} {\bibfnamefont {H.-Y.}\ \bibnamefont {Sun}},\
  }\bibfield  {title} {\bibinfo {title} {{Universal relations for holographic
  interfaces}},\ }\href {https://doi.org/10.1007/JHEP09(2021)172} {\bibfield
  {journal} {\bibinfo  {journal} {JHEP}\ }\textbf {\bibinfo {volume} {09}},\
  \bibinfo {pages} {172}},\ \Eprint {https://arxiv.org/abs/2107.02165}
  {arXiv:2107.02165 [hep-th]} \BibitemShut {NoStop}%
\bibitem [{\citenamefont {Karch}\ and\ \citenamefont
  {Wang}(2023)}]{Karch:2022vot}%
  \BibitemOpen
  \bibfield  {author} {\bibinfo {author} {\bibfnamefont {A.}~\bibnamefont
  {Karch}}\ and\ \bibinfo {author} {\bibfnamefont {M.}~\bibnamefont {Wang}},\
  }\bibfield  {title} {\bibinfo {title} {{Universal behavior of entanglement
  entropies in interface CFTs from general holographic spacetimes}},\ }\href
  {https://doi.org/10.1007/JHEP06(2023)145} {\bibfield  {journal} {\bibinfo
  {journal} {JHEP}\ }\textbf {\bibinfo {volume} {06}},\ \bibinfo {pages}
  {145}},\ \Eprint {https://arxiv.org/abs/2211.09148} {arXiv:2211.09148
  [hep-th]} \BibitemShut {NoStop}%
\bibitem [{\citenamefont {Gaiotto}\ and\ \citenamefont
  {Witten}(2009)}]{Gaiotto:2008sa}%
  \BibitemOpen
  \bibfield  {author} {\bibinfo {author} {\bibfnamefont {D.}~\bibnamefont
  {Gaiotto}}\ and\ \bibinfo {author} {\bibfnamefont {E.}~\bibnamefont
  {Witten}},\ }\bibfield  {title} {\bibinfo {title} {{Supersymmetric Boundary
  Conditions in N=4 Super Yang-Mills Theory}},\ }\href
  {https://doi.org/10.1007/s10955-009-9687-3} {\bibfield  {journal} {\bibinfo
  {journal} {J. Statist. Phys.}\ }\textbf {\bibinfo {volume} {135}},\ \bibinfo
  {pages} {789} (\bibinfo {year} {2009})},\ \Eprint
  {https://arxiv.org/abs/0804.2902} {arXiv:0804.2902 [hep-th]} \BibitemShut
  {NoStop}%
\bibitem [{\citenamefont {Gaiotto}\ and\ \citenamefont
  {Witten}(2010)}]{Gaiotto:2008sd}%
  \BibitemOpen
  \bibfield  {author} {\bibinfo {author} {\bibfnamefont {D.}~\bibnamefont
  {Gaiotto}}\ and\ \bibinfo {author} {\bibfnamefont {E.}~\bibnamefont
  {Witten}},\ }\bibfield  {title} {\bibinfo {title} {{Janus Configurations,
  Chern-Simons Couplings, And The theta-Angle in N=4 Super Yang-Mills
  Theory}},\ }\href {https://doi.org/10.1007/JHEP06(2010)097} {\bibfield
  {journal} {\bibinfo  {journal} {JHEP}\ }\textbf {\bibinfo {volume} {06}},\
  \bibinfo {pages} {097}},\ \Eprint {https://arxiv.org/abs/0804.2907}
  {arXiv:0804.2907 [hep-th]} \BibitemShut {NoStop}%
\bibitem [{\citenamefont {D'Hoker}\ \emph
  {et~al.}(2007{\natexlab{a}})\citenamefont {D'Hoker}, \citenamefont {Estes},\
  and\ \citenamefont {Gutperle}}]{DHoker:2007zhm}%
  \BibitemOpen
  \bibfield  {author} {\bibinfo {author} {\bibfnamefont {E.}~\bibnamefont
  {D'Hoker}}, \bibinfo {author} {\bibfnamefont {J.}~\bibnamefont {Estes}},\
  and\ \bibinfo {author} {\bibfnamefont {M.}~\bibnamefont {Gutperle}},\
  }\bibfield  {title} {\bibinfo {title} {{Exact half-BPS Type IIB interface
  solutions. I. Local solution and supersymmetric Janus}},\ }\href
  {https://doi.org/10.1088/1126-6708/2007/06/021} {\bibfield  {journal}
  {\bibinfo  {journal} {JHEP}\ }\textbf {\bibinfo {volume} {06}},\ \bibinfo
  {pages} {021}},\ \Eprint {https://arxiv.org/abs/0705.0022} {arXiv:0705.0022
  [hep-th]} \BibitemShut {NoStop}%
\bibitem [{\citenamefont {D'Hoker}\ \emph
  {et~al.}(2007{\natexlab{b}})\citenamefont {D'Hoker}, \citenamefont {Estes},\
  and\ \citenamefont {Gutperle}}]{DHoker:2007hhe}%
  \BibitemOpen
  \bibfield  {author} {\bibinfo {author} {\bibfnamefont {E.}~\bibnamefont
  {D'Hoker}}, \bibinfo {author} {\bibfnamefont {J.}~\bibnamefont {Estes}},\
  and\ \bibinfo {author} {\bibfnamefont {M.}~\bibnamefont {Gutperle}},\
  }\bibfield  {title} {\bibinfo {title} {{Exact half-BPS Type IIB interface
  solutions. II. Flux solutions and multi-Janus}},\ }\href
  {https://doi.org/10.1088/1126-6708/2007/06/022} {\bibfield  {journal}
  {\bibinfo  {journal} {JHEP}\ }\textbf {\bibinfo {volume} {06}},\ \bibinfo
  {pages} {022}},\ \Eprint {https://arxiv.org/abs/0705.0024} {arXiv:0705.0024
  [hep-th]} \BibitemShut {NoStop}%
\bibitem [{\citenamefont {Uhlemann}(2021)}]{Uhlemann:2021nhu}%
  \BibitemOpen
  \bibfield  {author} {\bibinfo {author} {\bibfnamefont {C.~F.}\ \bibnamefont
  {Uhlemann}},\ }\bibfield  {title} {\bibinfo {title} {{Islands and Page curves
  in 4d from Type IIB}},\ }\href {https://doi.org/10.1007/JHEP08(2021)104}
  {\bibfield  {journal} {\bibinfo  {journal} {JHEP}\ }\textbf {\bibinfo
  {volume} {08}},\ \bibinfo {pages} {104}},\ \Eprint
  {https://arxiv.org/abs/2105.00008} {arXiv:2105.00008 [hep-th]} \BibitemShut
  {NoStop}%
\bibitem [{\citenamefont {Bachas}\ and\ \citenamefont
  {Lavdas}(2018)}]{Bachas:2018zmb}%
  \BibitemOpen
  \bibfield  {author} {\bibinfo {author} {\bibfnamefont {C.}~\bibnamefont
  {Bachas}}\ and\ \bibinfo {author} {\bibfnamefont {I.}~\bibnamefont
  {Lavdas}},\ }\bibfield  {title} {\bibinfo {title} {{Massive Anti-de Sitter
  Gravity from String Theory}},\ }\href
  {https://doi.org/10.1007/JHEP11(2018)003} {\bibfield  {journal} {\bibinfo
  {journal} {JHEP}\ }\textbf {\bibinfo {volume} {11}},\ \bibinfo {pages}
  {003}},\ \Eprint {https://arxiv.org/abs/1807.00591} {arXiv:1807.00591
  [hep-th]} \BibitemShut {NoStop}%
\bibitem [{\citenamefont {De~Luca}\ \emph {et~al.}(2021)\citenamefont
  {De~Luca}, \citenamefont {De~Ponti}, \citenamefont {Mondino},\ and\
  \citenamefont {Tomasiello}}]{DeLuca:2021ojx}%
  \BibitemOpen
  \bibfield  {author} {\bibinfo {author} {\bibfnamefont {G.~B.}\ \bibnamefont
  {De~Luca}}, \bibinfo {author} {\bibfnamefont {N.}~\bibnamefont {De~Ponti}},
  \bibinfo {author} {\bibfnamefont {A.}~\bibnamefont {Mondino}},\ and\ \bibinfo
  {author} {\bibfnamefont {A.}~\bibnamefont {Tomasiello}},\ }\bibfield  {title}
  {\bibinfo {title} {{Cheeger bounds on spin-two fields}},\ }\href
  {https://doi.org/10.1007/JHEP12(2021)217} {\bibfield  {journal} {\bibinfo
  {journal} {JHEP}\ }\textbf {\bibinfo {volume} {12}},\ \bibinfo {pages}
  {217}},\ \Eprint {https://arxiv.org/abs/2109.11560} {arXiv:2109.11560
  [hep-th]} \BibitemShut {NoStop}%
\bibitem [{\citenamefont {Hanany}\ and\ \citenamefont
  {Witten}(1997)}]{Hanany:1996ie}%
  \BibitemOpen
  \bibfield  {author} {\bibinfo {author} {\bibfnamefont {A.}~\bibnamefont
  {Hanany}}\ and\ \bibinfo {author} {\bibfnamefont {E.}~\bibnamefont
  {Witten}},\ }\bibfield  {title} {\bibinfo {title} {{Type IIB superstrings,
  BPS monopoles, and three-dimensional gauge dynamics}},\ }\href
  {https://doi.org/10.1016/S0550-3213(97)00157-0} {\bibfield  {journal}
  {\bibinfo  {journal} {Nucl. Phys. B}\ }\textbf {\bibinfo {volume} {492}},\
  \bibinfo {pages} {152} (\bibinfo {year} {1997})},\ \Eprint
  {https://arxiv.org/abs/hep-th/9611230} {arXiv:hep-th/9611230} \BibitemShut
  {NoStop}%
\bibitem [{\citenamefont {Aharony}\ \emph {et~al.}(2011)\citenamefont
  {Aharony}, \citenamefont {Berdichevsky}, \citenamefont {Berkooz},\ and\
  \citenamefont {Shamir}}]{Aharony:2011yc}%
  \BibitemOpen
  \bibfield  {author} {\bibinfo {author} {\bibfnamefont {O.}~\bibnamefont
  {Aharony}}, \bibinfo {author} {\bibfnamefont {L.}~\bibnamefont
  {Berdichevsky}}, \bibinfo {author} {\bibfnamefont {M.}~\bibnamefont
  {Berkooz}},\ and\ \bibinfo {author} {\bibfnamefont {I.}~\bibnamefont
  {Shamir}},\ }\bibfield  {title} {\bibinfo {title} {{Near-horizon solutions
  for D3-branes ending on 5-branes}},\ }\href
  {https://doi.org/10.1103/PhysRevD.84.126003} {\bibfield  {journal} {\bibinfo
  {journal} {Phys. Rev. D}\ }\textbf {\bibinfo {volume} {84}},\ \bibinfo
  {pages} {126003} (\bibinfo {year} {2011})},\ \Eprint
  {https://arxiv.org/abs/1106.1870} {arXiv:1106.1870 [hep-th]} \BibitemShut
  {NoStop}%
\bibitem [{\citenamefont {Assel}\ \emph {et~al.}(2011)\citenamefont {Assel},
  \citenamefont {Bachas}, \citenamefont {Estes},\ and\ \citenamefont
  {Gomis}}]{Assel:2011xz}%
  \BibitemOpen
  \bibfield  {author} {\bibinfo {author} {\bibfnamefont {B.}~\bibnamefont
  {Assel}}, \bibinfo {author} {\bibfnamefont {C.}~\bibnamefont {Bachas}},
  \bibinfo {author} {\bibfnamefont {J.}~\bibnamefont {Estes}},\ and\ \bibinfo
  {author} {\bibfnamefont {J.}~\bibnamefont {Gomis}},\ }\bibfield  {title}
  {\bibinfo {title} {{Holographic Duals of D=3 N=4 Superconformal Field
  Theories}},\ }\href {https://doi.org/10.1007/JHEP08(2011)087} {\bibfield
  {journal} {\bibinfo  {journal} {JHEP}\ }\textbf {\bibinfo {volume} {08}},\
  \bibinfo {pages} {087}},\ \Eprint {https://arxiv.org/abs/1106.4253}
  {arXiv:1106.4253 [hep-th]} \BibitemShut {NoStop}%
\bibitem [{\citenamefont {Hubeny}\ \emph {et~al.}(2007)\citenamefont {Hubeny},
  \citenamefont {Rangamani},\ and\ \citenamefont {Takayanagi}}]{Hubeny:2007xt}%
  \BibitemOpen
  \bibfield  {author} {\bibinfo {author} {\bibfnamefont {V.~E.}\ \bibnamefont
  {Hubeny}}, \bibinfo {author} {\bibfnamefont {M.}~\bibnamefont {Rangamani}},\
  and\ \bibinfo {author} {\bibfnamefont {T.}~\bibnamefont {Takayanagi}},\
  }\bibfield  {title} {\bibinfo {title} {{A Covariant holographic entanglement
  entropy proposal}},\ }\href {https://doi.org/10.1088/1126-6708/2007/07/062}
  {\bibfield  {journal} {\bibinfo  {journal} {JHEP}\ }\textbf {\bibinfo
  {volume} {07}},\ \bibinfo {pages} {062}},\ \Eprint
  {https://arxiv.org/abs/0705.0016} {arXiv:0705.0016 [hep-th]} \BibitemShut
  {NoStop}%
\bibitem [{\citenamefont {Raamsdonk}\ and\ \citenamefont
  {Waddell}(2021)}]{Raamsdonk:2020tin}%
  \BibitemOpen
  \bibfield  {author} {\bibinfo {author} {\bibfnamefont {M.~V.}\ \bibnamefont
  {Raamsdonk}}\ and\ \bibinfo {author} {\bibfnamefont {C.}~\bibnamefont
  {Waddell}},\ }\bibfield  {title} {\bibinfo {title} {{Holographic and
  localization calculations of boundary F for $ \mathcal{N} $ = 4 SUSY
  Yang-Mills theory}},\ }\href {https://doi.org/10.1007/JHEP02(2021)222}
  {\bibfield  {journal} {\bibinfo  {journal} {JHEP}\ }\textbf {\bibinfo
  {volume} {02}},\ \bibinfo {pages} {222}},\ \Eprint
  {https://arxiv.org/abs/2010.14520} {arXiv:2010.14520 [hep-th]} \BibitemShut
  {NoStop}%
\bibitem [{\citenamefont {Bak}\ \emph {et~al.}(2003)\citenamefont {Bak},
  \citenamefont {Gutperle},\ and\ \citenamefont {Hirano}}]{Bak:2003jk}%
  \BibitemOpen
  \bibfield  {author} {\bibinfo {author} {\bibfnamefont {D.}~\bibnamefont
  {Bak}}, \bibinfo {author} {\bibfnamefont {M.}~\bibnamefont {Gutperle}},\ and\
  \bibinfo {author} {\bibfnamefont {S.}~\bibnamefont {Hirano}},\ }\bibfield
  {title} {\bibinfo {title} {{A Dilatonic deformation of AdS(5) and its field
  theory dual}},\ }\href {https://doi.org/10.1088/1126-6708/2003/05/072}
  {\bibfield  {journal} {\bibinfo  {journal} {JHEP}\ }\textbf {\bibinfo
  {volume} {05}},\ \bibinfo {pages} {072}},\ \Eprint
  {https://arxiv.org/abs/hep-th/0304129} {arXiv:hep-th/0304129} \BibitemShut
  {NoStop}%
\bibitem [{\citenamefont {Clark}\ \emph {et~al.}(2005)\citenamefont {Clark},
  \citenamefont {Freedman}, \citenamefont {Karch},\ and\ \citenamefont
  {Schnabl}}]{Clark:2004sb}%
  \BibitemOpen
  \bibfield  {author} {\bibinfo {author} {\bibfnamefont {A.~B.}\ \bibnamefont
  {Clark}}, \bibinfo {author} {\bibfnamefont {D.~Z.}\ \bibnamefont {Freedman}},
  \bibinfo {author} {\bibfnamefont {A.}~\bibnamefont {Karch}},\ and\ \bibinfo
  {author} {\bibfnamefont {M.}~\bibnamefont {Schnabl}},\ }\bibfield  {title}
  {\bibinfo {title} {{Dual of the Janus solution: An interface conformal field
  theory}},\ }\href {https://doi.org/10.1103/PhysRevD.71.066003} {\bibfield
  {journal} {\bibinfo  {journal} {Phys. Rev. D}\ }\textbf {\bibinfo {volume}
  {71}},\ \bibinfo {pages} {066003} (\bibinfo {year} {2005})},\ \Eprint
  {https://arxiv.org/abs/hep-th/0407073} {arXiv:hep-th/0407073} \BibitemShut
  {NoStop}%
\bibitem [{\citenamefont {D'Hoker}\ \emph {et~al.}(2006)\citenamefont
  {D'Hoker}, \citenamefont {Estes},\ and\ \citenamefont
  {Gutperle}}]{DHoker:2006qeo}%
  \BibitemOpen
  \bibfield  {author} {\bibinfo {author} {\bibfnamefont {E.}~\bibnamefont
  {D'Hoker}}, \bibinfo {author} {\bibfnamefont {J.}~\bibnamefont {Estes}},\
  and\ \bibinfo {author} {\bibfnamefont {M.}~\bibnamefont {Gutperle}},\
  }\bibfield  {title} {\bibinfo {title} {{Interface Yang-Mills, supersymmetry,
  and Janus}},\ }\href {https://doi.org/10.1016/j.nuclphysb.2006.07.001}
  {\bibfield  {journal} {\bibinfo  {journal} {Nucl. Phys. B}\ }\textbf
  {\bibinfo {volume} {753}},\ \bibinfo {pages} {16} (\bibinfo {year} {2006})},\
  \Eprint {https://arxiv.org/abs/hep-th/0603013} {arXiv:hep-th/0603013}
  \BibitemShut {NoStop}%
\bibitem [{\citenamefont {Coccia}\ and\ \citenamefont
  {Uhlemann}(2021)}]{Coccia:2020wtk}%
  \BibitemOpen
  \bibfield  {author} {\bibinfo {author} {\bibfnamefont {L.}~\bibnamefont
  {Coccia}}\ and\ \bibinfo {author} {\bibfnamefont {C.~F.}\ \bibnamefont
  {Uhlemann}},\ }\bibfield  {title} {\bibinfo {title} {{On the planar limit of
  3d $
  {\mathrm{T}}_{\rho}^{\sigma}\left[\mathrm{SU}\left(\mathrm{N}\right)\right]
  $}},\ }\href {https://doi.org/10.1007/JHEP06(2021)038} {\bibfield  {journal}
  {\bibinfo  {journal} {JHEP}\ }\textbf {\bibinfo {volume} {06}},\ \bibinfo
  {pages} {038}},\ \Eprint {https://arxiv.org/abs/2011.10050} {arXiv:2011.10050
  [hep-th]} \BibitemShut {NoStop}%
\bibitem [{\citenamefont {Coccia}\ and\ \citenamefont
  {Uhlemann}(2022)}]{Coccia:2021lpp}%
  \BibitemOpen
  \bibfield  {author} {\bibinfo {author} {\bibfnamefont {L.}~\bibnamefont
  {Coccia}}\ and\ \bibinfo {author} {\bibfnamefont {C.~F.}\ \bibnamefont
  {Uhlemann}},\ }\bibfield  {title} {\bibinfo {title} {{Mapping out the
  internal space in AdS/BCFT with Wilson loops}},\ }\href
  {https://doi.org/10.1007/JHEP03(2022)127} {\bibfield  {journal} {\bibinfo
  {journal} {JHEP}\ }\textbf {\bibinfo {volume} {03}},\ \bibinfo {pages}
  {127}},\ \Eprint {https://arxiv.org/abs/2112.14648} {arXiv:2112.14648
  [hep-th]} \BibitemShut {NoStop}%
\bibitem [{\citenamefont {Karch}\ \emph {et~al.}(2022)\citenamefont {Karch},
  \citenamefont {Sun},\ and\ \citenamefont {Uhlemann}}]{Karch:2022rvr}%
  \BibitemOpen
  \bibfield  {author} {\bibinfo {author} {\bibfnamefont {A.}~\bibnamefont
  {Karch}}, \bibinfo {author} {\bibfnamefont {H.}~\bibnamefont {Sun}},\ and\
  \bibinfo {author} {\bibfnamefont {C.~F.}\ \bibnamefont {Uhlemann}},\
  }\bibfield  {title} {\bibinfo {title} {{Double holography in string
  theory}},\ }\href {https://doi.org/10.1007/JHEP10(2022)012} {\bibfield
  {journal} {\bibinfo  {journal} {JHEP}\ }\textbf {\bibinfo {volume} {10}},\
  \bibinfo {pages} {012}},\ \Eprint {https://arxiv.org/abs/2206.11292}
  {arXiv:2206.11292 [hep-th]} \BibitemShut {NoStop}%
\bibitem [{\citenamefont {Bachas}\ \emph {et~al.}(2020)\citenamefont {Bachas},
  \citenamefont {Chapman}, \citenamefont {Ge},\ and\ \citenamefont
  {Policastro}}]{Bachas:2020yxv}%
  \BibitemOpen
  \bibfield  {author} {\bibinfo {author} {\bibfnamefont {C.}~\bibnamefont
  {Bachas}}, \bibinfo {author} {\bibfnamefont {S.}~\bibnamefont {Chapman}},
  \bibinfo {author} {\bibfnamefont {D.}~\bibnamefont {Ge}},\ and\ \bibinfo
  {author} {\bibfnamefont {G.}~\bibnamefont {Policastro}},\ }\bibfield  {title}
  {\bibinfo {title} {{Energy Reflection and Transmission at 2D Holographic
  Interfaces}},\ }\href {https://doi.org/10.1103/PhysRevLett.125.231602}
  {\bibfield  {journal} {\bibinfo  {journal} {Phys. Rev. Lett.}\ }\textbf
  {\bibinfo {volume} {125}},\ \bibinfo {pages} {231602} (\bibinfo {year}
  {2020})},\ \Eprint {https://arxiv.org/abs/2006.11333} {arXiv:2006.11333
  [hep-th]} \BibitemShut {NoStop}%
\bibitem [{\citenamefont {Bachas}\ \emph {et~al.}(2023)\citenamefont {Bachas},
  \citenamefont {Baiguera}, \citenamefont {Chapman}, \citenamefont
  {Policastro},\ and\ \citenamefont {Schwartzman}}]{Bachas:2022etu}%
  \BibitemOpen
  \bibfield  {author} {\bibinfo {author} {\bibfnamefont {C.}~\bibnamefont
  {Bachas}}, \bibinfo {author} {\bibfnamefont {S.}~\bibnamefont {Baiguera}},
  \bibinfo {author} {\bibfnamefont {S.}~\bibnamefont {Chapman}}, \bibinfo
  {author} {\bibfnamefont {G.}~\bibnamefont {Policastro}},\ and\ \bibinfo
  {author} {\bibfnamefont {T.}~\bibnamefont {Schwartzman}},\ }\bibfield
  {title} {\bibinfo {title} {{Energy Transport for Thick Holographic Branes}},\
  }\href {https://doi.org/10.1103/PhysRevLett.131.021601} {\bibfield  {journal}
  {\bibinfo  {journal} {Phys. Rev. Lett.}\ }\textbf {\bibinfo {volume} {131}},\
  \bibinfo {pages} {021601} (\bibinfo {year} {2023})},\ \Eprint
  {https://arxiv.org/abs/2212.14058} {arXiv:2212.14058 [hep-th]} \BibitemShut
  {NoStop}%
\bibitem [{\citenamefont {Ryu}\ and\ \citenamefont
  {Takayanagi}(2006)}]{Ryu:2006ef}%
  \BibitemOpen
  \bibfield  {author} {\bibinfo {author} {\bibfnamefont {S.}~\bibnamefont
  {Ryu}}\ and\ \bibinfo {author} {\bibfnamefont {T.}~\bibnamefont
  {Takayanagi}},\ }\bibfield  {title} {\bibinfo {title} {{Aspects of
  Holographic Entanglement Entropy}},\ }\href
  {https://doi.org/10.1088/1126-6708/2006/08/045} {\bibfield  {journal}
  {\bibinfo  {journal} {JHEP}\ }\textbf {\bibinfo {volume} {08}},\ \bibinfo
  {pages} {045}},\ \Eprint {https://arxiv.org/abs/hep-th/0605073}
  {arXiv:hep-th/0605073} \BibitemShut {NoStop}%
\bibitem [{\citenamefont {Solodukhin}(2008)}]{Solodukhin:2008dh}%
  \BibitemOpen
  \bibfield  {author} {\bibinfo {author} {\bibfnamefont {S.~N.}\ \bibnamefont
  {Solodukhin}},\ }\bibfield  {title} {\bibinfo {title} {{Entanglement entropy,
  conformal invariance and extrinsic geometry}},\ }\href
  {https://doi.org/10.1016/j.physletb.2008.05.071} {\bibfield  {journal}
  {\bibinfo  {journal} {Phys. Lett. B}\ }\textbf {\bibinfo {volume} {665}},\
  \bibinfo {pages} {305} (\bibinfo {year} {2008})},\ \Eprint
  {https://arxiv.org/abs/0802.3117} {arXiv:0802.3117 [hep-th]} \BibitemShut
  {NoStop}%
\bibitem [{\citenamefont {Graham}\ and\ \citenamefont
  {Witten}(1999)}]{Graham:1999pm}%
  \BibitemOpen
  \bibfield  {author} {\bibinfo {author} {\bibfnamefont {C.~R.}\ \bibnamefont
  {Graham}}\ and\ \bibinfo {author} {\bibfnamefont {E.}~\bibnamefont
  {Witten}},\ }\bibfield  {title} {\bibinfo {title} {{Conformal anomaly of
  submanifold observables in AdS / CFT correspondence}},\ }\href
  {https://doi.org/10.1016/S0550-3213(99)00055-3} {\bibfield  {journal}
  {\bibinfo  {journal} {Nucl. Phys. B}\ }\textbf {\bibinfo {volume} {546}},\
  \bibinfo {pages} {52} (\bibinfo {year} {1999})},\ \Eprint
  {https://arxiv.org/abs/hep-th/9901021} {arXiv:hep-th/9901021} \BibitemShut
  {NoStop}%
\bibitem [{\citenamefont {Skenderis}(2002)}]{Skenderis:2002wp}%
  \BibitemOpen
  \bibfield  {author} {\bibinfo {author} {\bibfnamefont {K.}~\bibnamefont
  {Skenderis}},\ }\bibfield  {title} {\bibinfo {title} {{Lecture notes on
  holographic renormalization}},\ }\href
  {https://doi.org/10.1088/0264-9381/19/22/306} {\bibfield  {journal} {\bibinfo
   {journal} {Class. Quant. Grav.}\ }\textbf {\bibinfo {volume} {19}},\
  \bibinfo {pages} {5849} (\bibinfo {year} {2002})},\ \Eprint
  {https://arxiv.org/abs/hep-th/0209067} {arXiv:hep-th/0209067} \BibitemShut
  {NoStop}%
\end{thebibliography}%
\end{document}